\documentclass[11pt]{article}

\usepackage{amsfonts,amssymb,chet,dsfont,graphicx,mathrsfs,pgfplots,tikz}

\usetikzlibrary{external}
\title{Anarchy and Neutrino Physics}

\author{Jean-Fran\c{c}ois Fortin\email{jean-francois.fortin@phy.ulaval.ca}, Nicolas Giasson\email{nicolas.giasson.1@ulaval.ca} and Luc Marleau\email{luc.marleau@phy.ulaval.ca}}

\affiliation{
D\'epartement de Physique, de G\'enie Physique et d'Optique,\\Universit\'e Laval, Qu\'ebec, QC G1V 0A6, Canada
}

\abstract{%
The neutrino sector of a seesaw-extended Standard Model is investigated under the anarchy hypothesis.  The previously derived probability density functions for neutrino masses and mixings, which characterize the type I-III seesaw ensemble of $N\times N$ complex random matrices, are used to extract information on the relevant physical parameters.  For $N=2$ and $N=3$, the distributions of the light neutrino masses, as well as the mixing angles and phases, are obtained using numerical integration methods.  A systematic comparison with the much simpler type II seesaw ensemble is also performed to point out the fundamental differences between the two ensembles.  It is found that the type I-III seesaw ensemble is better suited to accommodate experimental data.  Moreover, the results indicate a strong preference for the mass splitting associated to normal hierarchy.  However, since all permutations of the singular values are found to be equally probable for a particular mass splitting, predictions regarding the hierarchy of the mass spectrum remains out of reach in the framework of anarchy.
}

\date{February 2017} 

\begin{document}

\maketitle



\section{Introduction}\label{SIntro}

The neutrino sector of the Standard Model (SM) is quite peculiar.  Indeed, although the quark and charged lepton mass spectra are quite hierarchical, the neutrino spectrum is simple: all neutrinos are massless.  Neutrino oscillations \cite{Olive:2016xmw,Gonzalez-Garcia:2014bfa,Esteban:2016qun}, where neutrinos seemingly change flavor in flight, cannot be accommodated in the SM due to the masslessness of the neutrinos.  Neutrino oscillations thus imply massive neutrino eigenstates and the SM must be extended.  Moreover, neutrino oscillation experimental data suggest that the neutrino spectrum is not hierarchical, with three massive light neutrinos and a mixing matrix, the Pontecorvo-Maki-Nakagawa-Sakata (PMNS) matrix, exhibiting near-maximal mixing.

The anarchy principle, introduced in \cite{Hall:1999sn,Haba:2000be}, was put forward to explain the peculiarities of the neutrino sector by postulating a low-energy neutrino mass matrix generated by one of the seesaw mechanisms from randomly-generated high-energy mass matrices with elements distributed following a Gaussian ensemble.  In a series of papers \cite{deGouvea:2003xe,deGouvea:2012ac,Heeck:2012fw,Bai:2012zn,Lu:2014cla,Babu:2016aro}, several numerical analysis of the anarchy principle were performed by randomly generating high-energy mass matrices and computing the corresponding low-energy neutrino mass matrices.

Recently, the low-energy neutrino mass matrix probability density function (pdf) for the anarchy principle was obtained from first principles in \cite{Fortin:2016zyf} following the extensive literature on random matrix theory \cite{mehta2004random,muirhead2009aspects,forrester2010log}.  It was shown that the pdfs associated to type I and type III seesaw mechanisms were given by the same complicated integral equation while the pdfs associated to type II seesaw mechanism was simple.  A partial investigation of these seesaw ensembles was completed in \cite{Fortin:2016zyf} but an analysis of the physical case of three light neutrinos was not undertaken.  This paper closes the gap by studying the implications of the seesaw ensembles for neutrino physics.

The low-energy neutrino mass matrix pdf is known to factorize into a singular value pdf and a group variable pdf for all seesaw mechanisms.  The singular value pdf corresponds to the pdf for the light neutrino masses while the group variable pdf corresponds to the pdf for the mixing angles and phases of the PMNS matrix.  The singular value pdf for type I-III is given in \cite{Fortin:2016zyf} as a multidimensional integral while the singular value pdf for type II is a simple Gaussian-like distribution.  The group variable pdf for all types of seesaw mechanisms is the Haar measure, which seems to prefer near-maximal mixing.  However, as stressed in \cite{Espinosa:2003qz}, the mode of a pdf is not a well-defined quantity (\textit{e.g.} it is not invariant under change of variables), hence it is preferable to compare pdfs by comparing their probabilities associated to a particular outcome.  This probability test, which is hard to perform by randomly generating low-energy neutrino mass matrices, can however be straightforwardly implemented from the analytic pdfs.

Moreover, the factorization of the pdfs for the singular values and the group variables implies that there is no link between the light neutrino masses and the light neutrino mass eigenstates, forbidding an investigation of the preferred mass hierarchy (normal or inverted).  From the analytic results for the singular value pdfs, it is however possible to determine which mass splitting, \textit{i.e.} which ordering of $\hat{m}^{2}_{\text{med}}-\hat{m}^{2}_{\text{min}}$ and $\hat{m}^{2}_{\text{max}}-\hat{m}^{2}_{\text{med}}$, is favored.

This paper is organized as follows: Section \ref{SSeesaw} gives the relevant pdfs for the seesaw ensembles obtained in \cite{Fortin:2016zyf}.  In section \ref{SComplexSeesaw} the pdfs for the complex seesaw ensembles are studied in the $N=2$ and $N=3$ cases.  For both cases, a comparison is made between the analytic results and the numerical results, showing perfect agreement.  For the $N=3$ case relevant to neutrino physics, a thorough investigation of the implications of the seesaw ensembles is completed.  For example, it is shown from the probability test that both type I-III and type II seesaw ensembles prefer the mass splitting associated to normal hierarchy with a neutrino energy scale of $\mathcal{O}(10^{-2})\:\text{eV}$.  Finally, a discussion and a conclusion are presented in section \ref{SConclusion}.

It is important to note that throughout this paper, the term ``analytic results'' should be understood as results obtained from the analytic pdfs which are numerically integrated while the term ``numerical results'' corresponds to results obtained from randomly-generated mass matrices.


\section{The Seesaw Ensembles}\label{SSeesaw}

This section states without proof the relevant quantities that appear in the seesaw ensembles, which were derived from the anarchy principle applied to the SM extended with the type I-III seesaw mechanism or with the type II seesaw mechanism.  The reader interested in the proofs is referred to \cite{Fortin:2016zyf}.

\subsection{Probability Density Functions}\label{SsPDF}

The pdfs for the dimensionless $N\times N$ light neutrino mass matrix $\hat{M}_\nu=M_\nu/(\sqrt{2}\Lambda_\nu)$ where $\Lambda_{\nu}$ is the (naturally small) light neutrino mass scale, can be expressed in terms of the light neutrino mass matrix singular values $\hat{m}_{\nu,i}$ and the light neutrino mass matrix group variables $U_\nu$ with the help of the decomposition $\hat{M}_\nu=U_\nu D_\nu U_\nu^T$ where $D_\nu=\text{diag}(\hat{m}_{\nu,1},\cdots,\hat{m}_{\nu,N})$ and $\hat{m}_{\nu,i}\geq0$ for all $i$.  The pdfs are found to factorize into two independent pdfs, one pdf for the singular values and one pdf for the group variables, as in
\eqn{P_\nu(\hat{m}_\nu;U_\nu)d\hat{m}_\nu dU_\nu=P_\nu(\hat{m}_\nu)P_\nu(U_\nu)d\hat{m}_\nu dU_\nu.}
For real ($\beta=1$) and complex ($\beta=2$) matrix elements, the pdfs are given respectively by
\eqn{
\begin{gathered}
P_\nu^\text{I-III}(\hat{m}_\nu)d\hat{m}_\nu=C_{\nu N}^{\text{I-III}\beta}I_N^\beta(\hat{m}_{\nu,1},\cdots,\hat{m}_{\nu,N})\prod_{1\leq i<j\leq N}|\hat{m}_{\nu,i}^\beta-\hat{m}_{\nu,j}^\beta|\prod_{1\leq i\leq N}|\hat{m}_{\nu,i}|^{-(\beta N+1)}d\hat{m}_{\nu,i},\\
P_\nu^\text{II}(\hat{m}_\nu)d\hat{m}_\nu=C_{\nu N}^{\text{II}\beta}\prod_{1\leq i<j\leq N}|\hat{m}_{\nu,i}^\beta-\hat{m}_{\nu,j}^\beta|\prod_{1\leq i\leq N}|\hat{m}_{\nu,i}|^{\beta-1}e^{-\hat{m}_{\nu,i}^2}d\hat{m}_{\nu,i},\\
P_\nu(U_\nu)dU_\nu=\frac{U_\nu^\dagger dU_\nu}{\text{Vol}(\mathcal{V}_N^\beta)}.
\end{gathered}
}[EqnPDF]
Note that the pdf for the group variables is the normalized Haar measure for all types of seesaw mechanisms.  Here the function $I_N^\beta$ relevant for the type I-III seesaw mechanism is
\eqna{
I_N^\beta(t_1,\cdots,t_N)&=\int_{U\in\mathcal{V}_N^\beta}\int_0^\infty\prod_{1\leq i<j\leq N}|x_i-x_j|^\beta e^{-2|\sum_{1\leq k\leq N}t_k^{-1}U_{ki}U_{kj}|^2x_ix_j}\\
&\phantom{=}\hspace{0.5cm}\times\prod_{1\leq i\leq N}x_i^{\beta(N+2)/2-1}e^{-x_i(1+|\sum_{1\leq j\leq N}t_j^{-1}U_{ji}^2|^2x_i)}dx_i\frac{(U^\dagger dU)'}{\text{Vol}(\mathcal{V}_N^\beta)/(2\pi)^{(\beta-1)N}},
}[EqnI]
where the integration is over the full Stiefel manifold $\mathcal{V}_N^\beta\equiv\mathcal{V}_{N,N}^\beta$ if $\beta=1$ but only parts of the full Stiefel manifold if $\beta=2$ (hence the prime, see \cite{Fortin:2016zyf}).  The normalization constants and the volume of the Stiefel manifold $\mathcal{V}_N^\beta$ are
\eqna{
C_{\nu N}^{\text{I-III}\beta}&=\frac{2^{N[\beta(N+3)-4]/4}}{N!}\prod_{1\leq i\leq N}\frac{\Gamma(\beta/2+1)}{\Gamma(\beta i/2+1)[\Gamma(\beta i/2)]^2},\\
C_{\nu N}^{\text{II}\beta}&=\frac{2^{N[\beta(N+3)-4]/4}}{N!}\prod_{1\leq i\leq N}\frac{1}{\Gamma(\beta i/2)},\\
\text{Vol}(\mathcal{V}_N^\beta)&=\int_{U\in\mathcal{V}_N^\beta}U^\dagger dU=\frac{2^N\pi^{\beta N(N+1)/4}}{\prod_{1\leq i\leq N}\Gamma(\beta i/2)}.
}
An important feature of the joint pdfs for the singular values \eqref{EqnPDF} is their invariance under permutations of the singular values.  Apart from the function $I_N^\beta$, the pdfs are clearly invariant under such a transformation.  The function $I_N^\beta$ is also invariant under permutations since a permutation only reshuffles the columns and rows of the matrix $U$, which is integrated over.\footnote{The permutation of $U$ can also be absorbed in an appropriate permutation of the variables $x_i$, which are also integrated over.}  Hence, the neutrino masses can be reshuffled freely amongst themselves.  This invariance, which leads to a complete independence between the light neutrino mass eigenstates and the light neutrino masses, implies that the probability for a specific spectrum of masses is at most $1/N!$.  This observation has far-reaching consequences in the physical case of neutrino physics.

It is important to note that, at the level of the pdfs, the only difference between the type I-III and the type II seesaw mechanisms lies in the singular value pdfs.  Hence, when comparing which ensemble better generates the observed neutrino parameters, only the information on the mass splittings will differ.

\subsection{A Parametrization for Unitary Matrices}\label{SsU}

Since the goal of this paper is to compare the implications of the seesaw ensembles with actual neutrino observations, the rest of the paper focuses on complex matrix elements, \textit{i.e.} $\beta=2$.  The analysis of the case with real matrix elements is similar.

With that in mind, it is important to find a convenient parametrization for unitary matrices to proceed with the analysis.  Indeed, to write the integral \eqref{EqnI} more explicitly when $\beta=2$, it is necessary to assume a parametrization for the unitary matrix $U$.  Moreover, the light neutrino group variables pdf is the Haar measure for unitary matrices.

A convenient parametrization, based on \cite{1751-8121-43-38-385306,:/content/aip/journal/jmp/53/1/10.1063/1.3672064}, implies that a $N\times N$ unitary matrix $U$ can be written as
\eqn{U=\prod_{1\leq j<k\leq N}\exp(i\phi_{jk}P_k)\exp(i\theta_{jk}\Sigma_{jk})\prod_{1\leq j\leq N}\exp(i\varphi_jP_j).}[EqnU]
Here the matrices $P_j$ and $\Sigma_{jk}$ are given by
\eqn{P_{j,ik}=\delta_{ji}\delta_{jk},\quad\quad\Sigma_{jk,i\ell}=-i\delta_{ji}\delta_{k\ell}+i\delta_{j\ell}\delta_{ki},}
and the $N^2$ mixing angles $\theta_{jk}$ and phases $\phi_{jk}$ and $\varphi_j$ belong to the following intervals,
\eqn{\theta_{jk}\in[0,\pi/2),\quad\quad\phi_{jk}\in[0,2\pi),\quad\quad\varphi_j\in[0,2\pi).}
Moreover, the Haar measure
\eqn{U^\dagger dU=\prod_{1\leq i\leq N}d\varphi_i\prod_{1\leq i<j\leq N}\sin(\theta_{ij})[\cos(\theta_{ij})]^{2(j-i)-1}d\phi_{ij}d\theta_{ij},}[EqnHaar]
depends only on the mixing angles.

For the light neutrino masses, the parametrization \eqref{EqnU}, which gives
\eqn{U^\dagger dU=U^\dagger d\left[\prod_{1\leq j<k\leq N}\exp(i\phi_{jk}P_k)\exp(i\theta_{jk}\Sigma_{jk})\right]\prod_{1\leq j\leq N}\exp(i\varphi_jP_j)+\text{diag}(id\varphi_1,\cdots,id\varphi_N),}
shows clearly that all $d\varphi_i$ are not integrated over in \eqref{EqnI}.  Indeed, the most interesting quantity in \eqref{EqnI} is $(U^\dagger dU)'$ and it is given by
\eqn{(U^\dagger dU)'=\bigwedge_{1\leq i<j\leq N}(U^\dagger dU)_{ij}=\bigwedge_{1\leq i<j\leq N}\text{Re}(U^\dagger dU)_{ij}\wedge\text{Im}(U^\dagger dU)_{ij},}
using the wedge product notation \cite{Fortin:2016zyf}.  Hence the variables $\varphi_i$ are not even part of the integral, implying a $N^2$-dimensional integral $I_N^{\beta=2}$ in the singular value pdf.

For the light neutrino mixing matrix, the quantity of interest is simply the Haar measure \eqref{EqnHaar}.  In that case, the corresponding variables $\varphi_i$ are the unphysical phases that can be absorbed by a field redefinition.  The remaining phases $\phi_{ij}$ are the CP-violating Dirac and Majorana phases.  All those phases have flat distributions that are uninteresting.  The mixing angles on the other hand, have non-trivial distributions.  This fact has important consequences for the SM neutrino physics since near-maximal mixings seem highly probable.


\section{The Complex Seesaw Ensembles}\label{SComplexSeesaw}

In this section the complex seesaw ensemble pdfs \eqref{EqnPDF} are analyzed for small values of $N$.  The case $N=1$ was studied analytically in \cite{Fortin:2016zyf} for both real and complex matrix elements.  Here the case $N=2$ with complex matrix elements is investigated and compared to numerical results for $2\times2$ matrices.  Then the physical case of $N=3$ is analyzed to determine how likely the pdfs are to generate the observed light neutrino masses and mixings.

\subsection{Consequences of the Complex Seesaw Ensembles}\label{SsConsequences}

Before discussing specific values of $N$, it is enlightening to state the implications of the complex seesaw ensemble pdfs \eqref{EqnPDF} in general terms.

First, once a decomposition for the light neutrino mass matrix $\hat{M}_\nu=U_\nu D_\nu U_\nu^T$ is chosen, the implications of near-maximal mixings obtained from the group variable pdf \eqref{EqnPDF}, \textit{i.e} the appropriate normalized Haar measure, for some mixing angles seem inescapable.  Indeed, using the parameterization \eqref{EqnU}, the Haar measure, given by \eqref{EqnHaar}, dictates that the most probable value for the mixing angles $\theta_{ij}$ is $\text{arccot}[\sqrt{2(j-i)-1}]$ while all the remaining (unphysical, CP-violating Dirac and Majorana) phases have flat distributions.  Thus the preferred value for all mixing angles $\theta_{i,i+1}$ is $\pi/4$, which corresponds to maximal mixing, while the preferred value for all mixing angles $\theta_{i,i+2}$ is $\pi/6$.  It is important however to notice that most probable values are not invariant under change of variables, as pointed out in \cite{Espinosa:2003qz}.

Then, the consequences for the light neutrino masses, which are obtained from the singular value pdfs \eqref{EqnPDF}, are not as sharp.  Indeed, although the chosen decomposition is here fixed, the singular value pdfs \eqref{EqnPDF} are invariant under permutations of the singular values.  Hence, the light neutrino masses and the light neutrino mixing angles and phases are completely independent.  In other words, although each singular value $\hat{m}_{\nu,i}$ has a corresponding singular vector $(\boldsymbol{u}_{\nu,i})_j=U_{\nu,ji}$ such that $\hat{M}_\nu\boldsymbol{u}_{\nu,i}^*=\hat{m}_{\nu,i}\boldsymbol{u}_{\nu,i}$, the probability that a particular neutrino mass spectrum occurs is at most $1/N!$.  For example, the dimensionless neutrino mass spectrum $(\hat{m}_{\nu,1},\cdots,\hat{m}_{\nu,N})=(\mu_1,\cdots,\mu_N)$ is as probable as the spectrum $(\hat{m}_{\nu,1},\cdots,\hat{m}_{\nu,N})=(\mu_2,\mu_1,\mu_3,\cdots,\mu_N)$ or any other permutations.  It is thus possible to fix an ordering for the singular values, $0\leq\hat{m}_\text{min}\leq\cdots\leq\hat{m}_\text{max}$, keeping in mind that the relationship between the light neutrino masses and the light neutrino mixing matrix is completely lost.  Fixing the ordering implies that the singular value pdfs \eqref{EqnPDF} are multiplied by $N!$.

Therefore, by working with a fixed basis as described above, some mixing angle preferred values correspond to maximal mixing but the ordering of the light neutrino masses for a given spectrum is completely free.  It is thus clear that a spectrum exhibiting one of the two hierarchy patterns preferred by the data (normal or inverse) is as probable as the same spectrum but with permuted mass eigenstates.  These observations are in the spirit of \cite{Espinosa:2003qz}, although with the analytic knowledge of the singular value pdfs \eqref{EqnPDF}, it is now possible to complete an appropriate statistical test to better check the validity of the anarchy principle.  As stressed in \cite{Espinosa:2003qz}, the most appropriate statistical test seems to be the probability test which computes the probability that the variables are in a given volume.  By choosing the observed values with their error bars for the volume of the light neutrino masses and mixings, the probability that one ensemble leads to the SM is obtained.  Clearly, since the variables are continuous, the calculated probability is very small for very precisely-known observed values.  One can nevertheless discriminate ensembles, for example the type I-III and type II seesaw ensembles, by comparing their respective probabilities, as shown below.

\subsection{The Case \texorpdfstring{$N=2$}{N=2}}\label{SsN2}

\begin{figure}[!t]
\centering
\resizebox{15cm}{!}{
\includegraphics{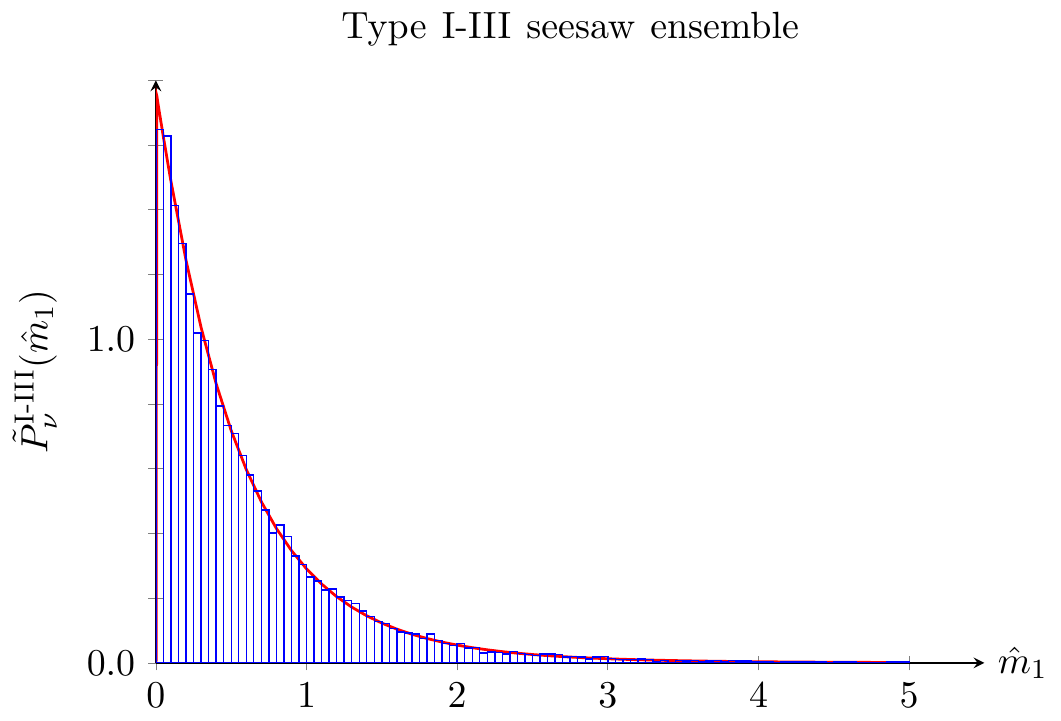}
\hspace{2cm}
\includegraphics{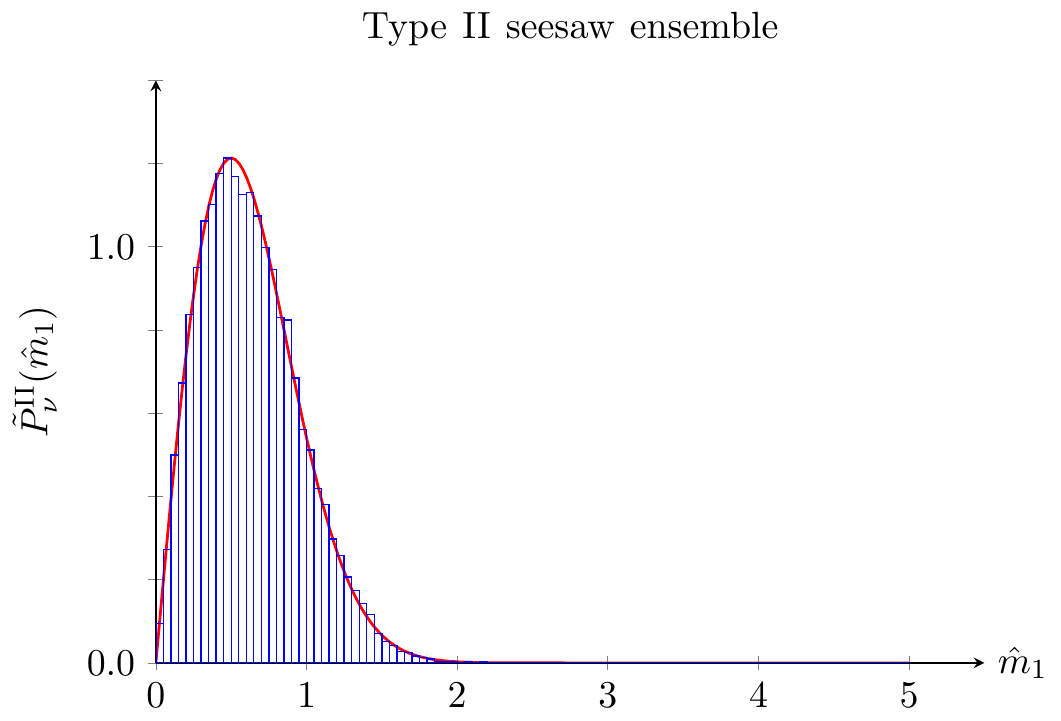}
}
\resizebox{15cm}{!}{
\includegraphics{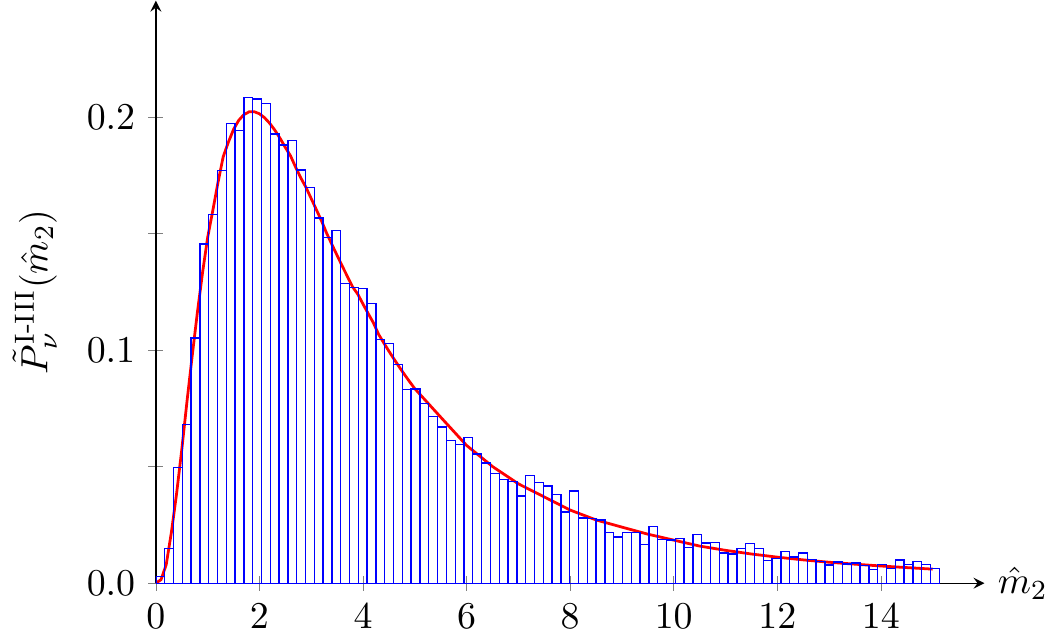}
\hspace{2cm}
\includegraphics{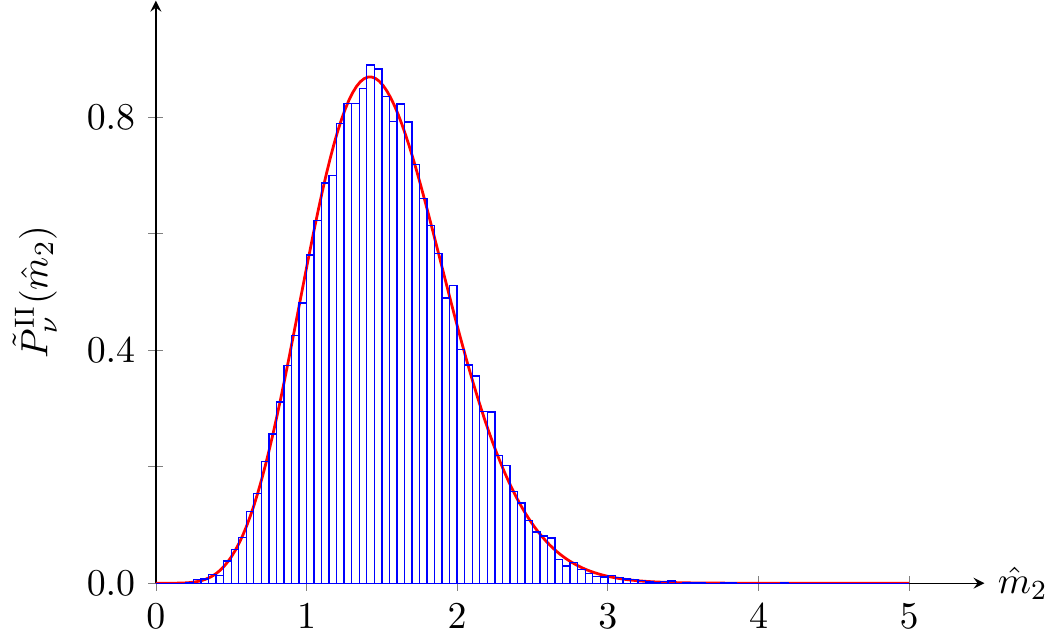}
}
\caption{Probability density functions for the singular values of the complex seesaw ensembles with $N=2$.  The red curve corresponds to the analytic result while the histogram corresponds to numerical results (with $2.5\times10^4$ dimensionless light neutrino mass matrices generated).  The left and right columns show the pdfs for the type I-III and the type II seesaw mechanisms respectively.  The singular values are ordered such that $0\leq\hat{m}_1\leq\hat{m}_2$ and an extra factor of $2!$ is introduced to correct the singular value pdfs.}
\label{FigPDFN2}
\end{figure}
\begin{figure}[!t]
\centering
\resizebox{15cm}{!}{
\includegraphics{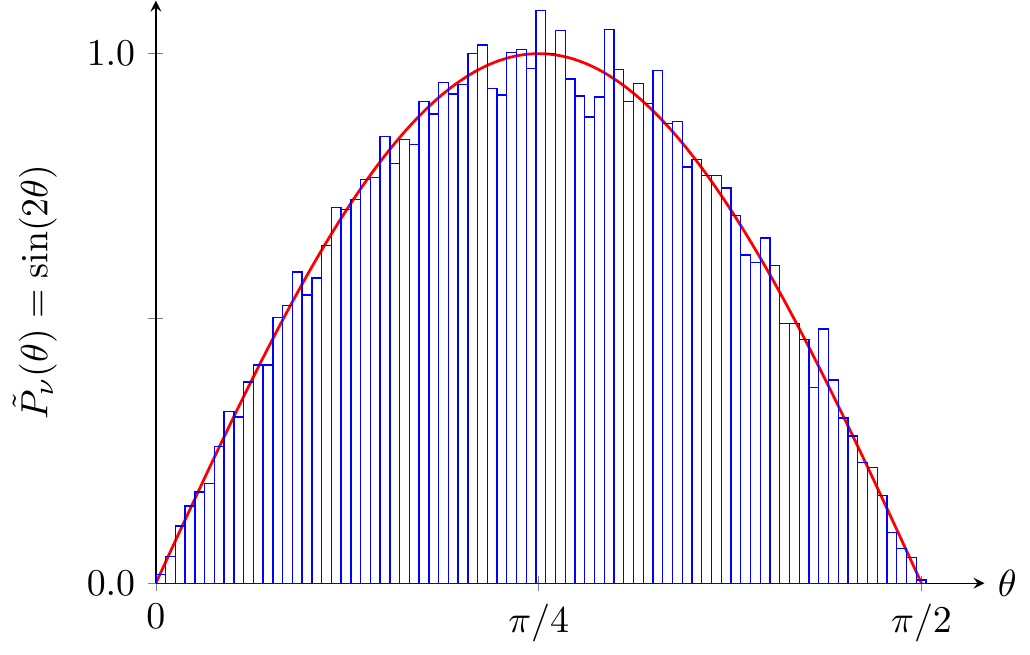}
\hspace{2cm}
\includegraphics{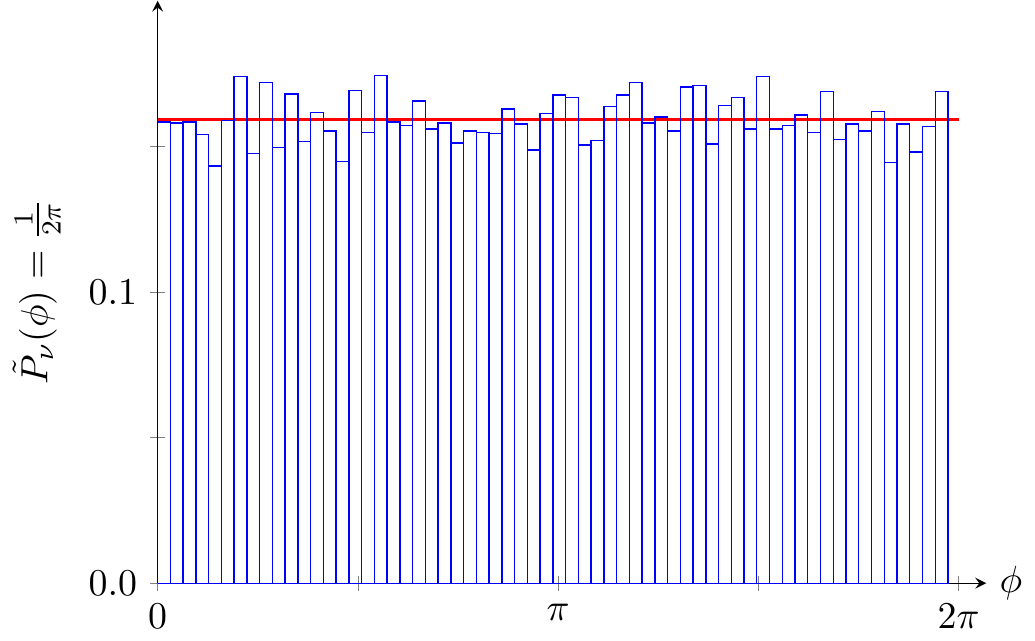}
}
\resizebox{15cm}{!}{
\includegraphics{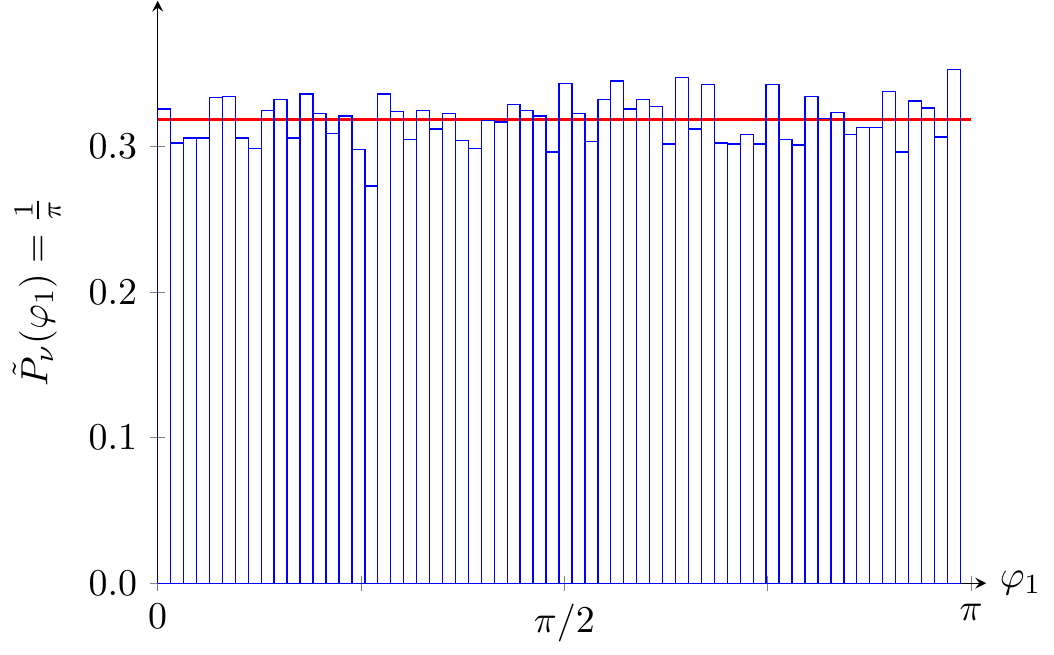}
\hspace{2cm}
\includegraphics{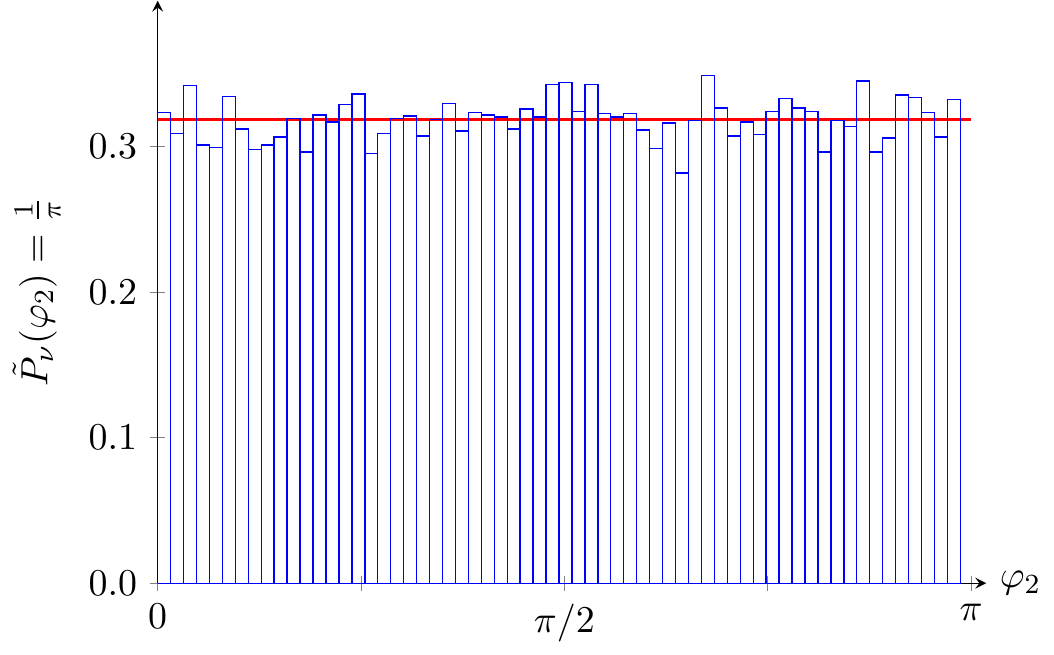}
}
\caption{Probability density functions for the mixing angle and phases of the complex seesaw ensembles with $N=2$.  The red curve corresponds to the analytic result while the histogram corresponds to numerical results (with $2.5\times10^4$ dimensionless light neutrino mass matrices generated).  The top and bottom rows show the pdfs for the mixing angle $\theta$, the CP-violating phase $\delta$, and the unphysical phases $\varphi_1$ and $\varphi_2$ (note the range is halved due to the extra freedom $\varphi_i\to\varphi_i\pm\pi$ \cite{Fortin:2016zyf}).}
\label{FigPDFN2ang}
\end{figure}

As a warm-up exercise, the case $N=2$ is studied analytically and compared to randomly-generated light neutrino mass matrices.  Using the parametrization \eqref{EqnU} for both unitary matrices [\textit{i.e.} the one appearing in \eqref{EqnI} and the light neutrino mixing matrix], the relevant pdfs \eqref{EqnPDF} can be written as
\eqn{
\begin{gathered}
P_\nu^\text{I-III}(\hat{m}_1,\hat{m}_2)=\frac{2|\hat{m}_1^2-\hat{m}_2^2|}{\hat{m}_1^5\hat{m}_2^5}\int_0^\infty dx_1dx_2\int_0^{\pi/2}d\theta'(x_1-x_2)^2x_1^3x_2^3\sin(2\theta')I_0\left[\frac{(x_1-x_2)^2[\sin(2\theta')]^2}{2\hat{m}_1\hat{m}_2}\right]\\
\hspace{2cm}\times e^{-\frac{\hat{m}_1^2(x_1[\sin(\theta')]^2+x_2[\cos(\theta')]^2)^2+\hat{m}_2^2(x_1[\cos(\theta')]^2+x_2[\sin(\theta')]^2)^2}{\hat{m}_1^2\hat{m}_2^2}-x_1-x_2},\\
P_\nu^\text{II}(\hat{m}_1,\hat{m}_2)=4|\hat{m}_1^2-\hat{m}_2^2|\hat{m}_1\hat{m}_2e^{-\hat{m}_1^2-\hat{m}_2^2},\\
P_\nu(\theta,\phi,\varphi_1,\varphi_2)=\frac{1}{8\pi^3}\sin(2\theta),
\end{gathered}
}[EqnPDFN2]
where the subscript $\nu$ on the masses was omitted to simplify the equations.
\begin{table}[!t]
\centering
\resizebox{8cm}{!}{%
\begin{tabular}{|c||c|c|c|}
\hline
Marginal pdfs & Mean & Median & Mode \\\hline
$\tilde{P}_\nu^\text{I-III}(\hat{m}_1)$ & $0.59$ & $0.39$ & $0.0064$\\
$\tilde{P}_\nu^\text{I-III}(\hat{m}_2)$ & $5.39$ & $3.41$ & $1.85$\\\hline
$\tilde{P}_\nu^\text{II}(\hat{m}_1)$ & $\frac{1}{2}\sqrt{\frac{\pi}{2}}$ & $0.58$ & $\frac{1}{2}$\\
$\tilde{P}_\nu^\text{II}(\hat{m}_2)$ & $\frac{\sqrt{\pi}}{4}\left(2+\sqrt{2}\right)$ & $1.48$ & $1.42$\\
\hline
\end{tabular}
}
\caption{Location parameters for the marginal singular value pdfs of figure \ref{FigPDFN2}.}
\label{TabparadistN2}
\end{table}
The modified Bessel function of the first kind $I_0(z)$ is generated by the integral over the phase in \eqref{EqnHaar}.  The $4$-dimensional integral is thus simplified to a $3$-dimensional integral.  Figures \ref{FigPDFN2} and \ref{FigPDFN2ang} show a comparison between the analytic results \eqref{EqnPDFN2} and numerical results for a fixed ordering of the singular values (chosen to be $0\leq\hat{m}_1\leq\hat{m}_2$, such that $\hat{m}_1\equiv\hat{m}_\text{min}$ and $\hat{m}_2\equiv\hat{m}_\text{max}$).  Consequently, the resulting marginal singular value pdfs are obtained by computing the following integrals
\eqn{\tilde{P}_\nu^\varSigma(\hat{m}_1)=2!\int_{\hat{m}_1}^\infty d\hat{m}_2P_\nu^\varSigma(\hat{m}_1,\hat{m}_2),\quad\quad\tilde{P}_\nu^\varSigma(\hat{m}_2)=2!\int_0^{\hat{m}_2}d\hat{m}_1P_\nu^\varSigma(\hat{m}_1,\hat{m}_2),}
for both ensembles (\textit{i.e.} $\varSigma=\text{I-III}$ or $\text{II}$).

Although the analytic behavior of the type I-III singular value pdf \eqref{EqnPDFN2} is hard to see intuitively due to the integral, it is clear that the pdfs \eqref{EqnPDFN2} are correct as seen in figure \ref{FigPDFN2}. The behavior of the type I-III singular value pdf at vanishing and large singular values matches the expectations of \cite{Fortin:2016zyf}.  The vanishing of $\tilde{P}_\nu^\text{I-III}(\hat{m}_1)$ at $\hat{m}_1\to0$ is not apparent in the histogram due to the bins being too large. The type II singular value pdf is much easier to study.  The pdfs for the smallest and the largest singular values can be obtained analytically from \eqref{EqnPDFN2}.  Again, there is a good match between the analytic results, which are simple exponentials, and the numerical results.  In order to make an appropriate comparison between the two ensembles, it becomes essential to fully characterize the previous pdfs.  Thus, one is expected to compute their corresponding moments as well as their modes and medians.  However, as stated in \cite{Fortin:2016zyf}, the only existing moment for the type I-III complex ($\beta=2$) seesaw ensemble is the first moment (the average singular values).  Therefore, a meaningful characterization of the previous pdfs is limited to the first moment, the mode and the median (their location parameters).  Their respective values for each distribution are presented in table \ref{TabparadistN2}.

From these results, it can be seen that the average singular values coming from the type I-III seesaw ensemble are spread over a wider range than in the type II seesaw ensemble.  Moreover, when comparing the mean of a distribution with its respective median, one can quantify the asymmetry of the pdfs presented in figure \ref{FigPDFN2}.  It turns out that the means are much closer to the medians (and thus the modes) in the type II seesaw ensemble, which leads to more symmetrical pdfs as can already be seen from figure \ref{FigPDFN2}. 

Moving forward, the group variable pdf \eqref{EqnPDFN2} is easier to analyze.  First, all phases have flat distributions as mentioned previously.  Moreover, the mixing angle has a non-trivial distribution that prefers near-maximal mixing.  From figure \ref{FigPDFN2ang} the pdf for the mixing angle and the phases agree well with the normalized Haar measure.  Since these pdfs were studied extensively in the literature and are easier to analyze, their statistical parameters (mean, median and mode), which are easily obtained from \eqref{EqnPDFN2}, are not presented here.

\begin{figure}[!t]
\centering
\resizebox{15cm}{!}{
\includegraphics{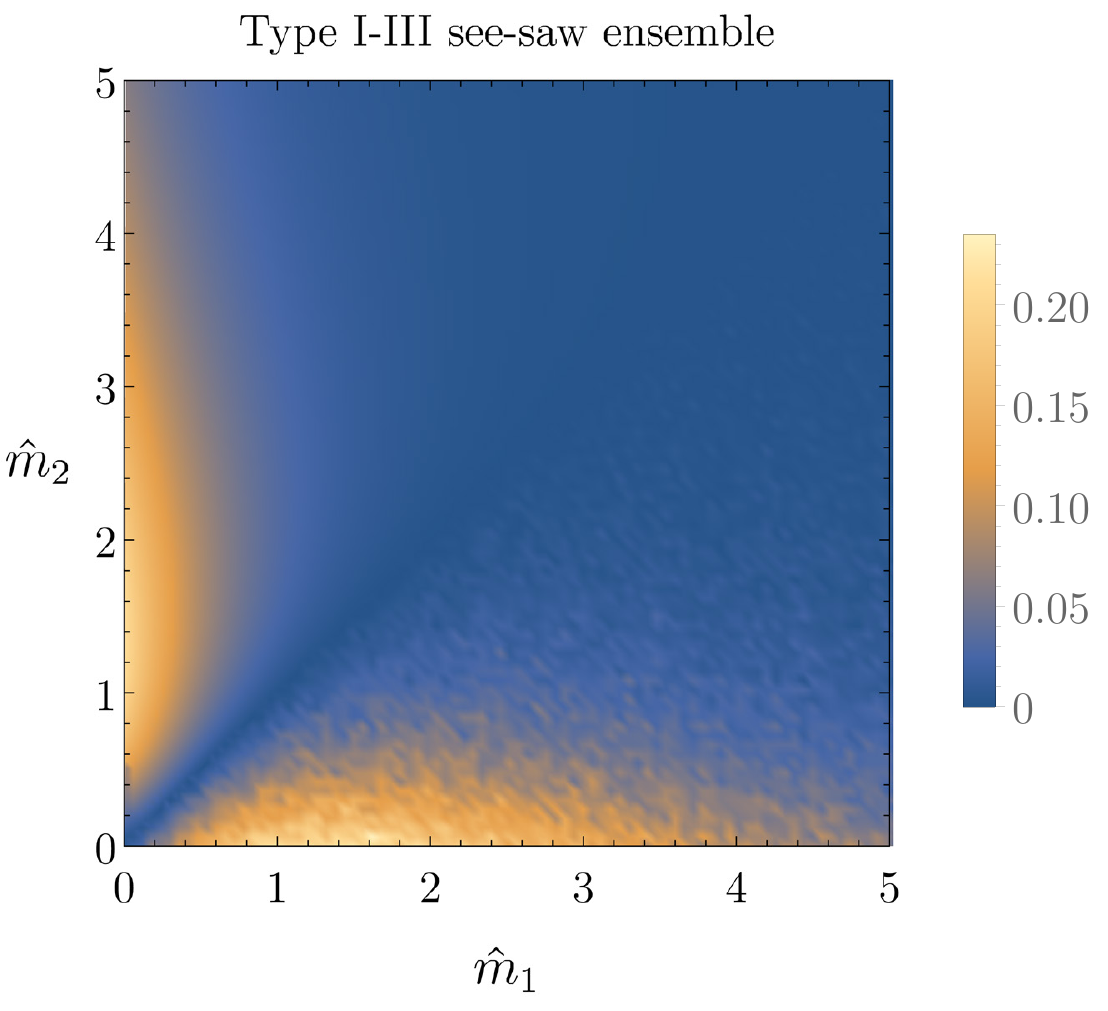}
\hspace{2cm}
\includegraphics{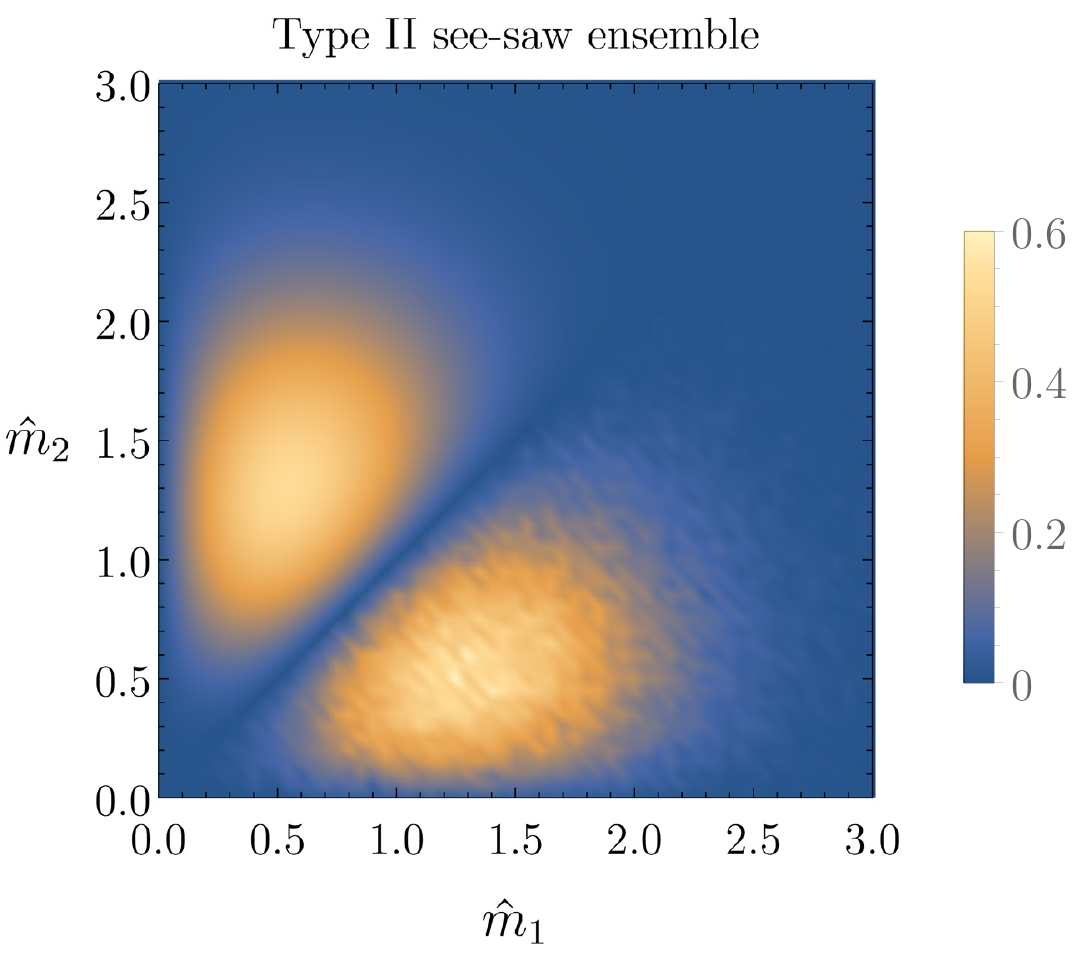}
}
\caption{Density plots of the singular value pdfs for the complex seesaw ensembles with $N=2$.  The left and right panels show the density plots of the type I-III $[P_\nu^\text{I-III}(\hat{m}_1,\hat{m}_2)]$ and type II $[P_\nu^\text{II}(\hat{m}_1,\hat{m}_2)]$ seesaw ensembles respectively.  The plots are split along the symmetry axis so that the upper and lower triangles show the analytical and numerical results (with $10^5$ dimensionless light neutrino mass matrices generated) respectively.  There is no ordering of the singular values.}
\label{FigPDFN2den}
\end{figure}

Finally, to provide a global overview of the pdfs with unordered singular values, the density plots of the singular value pdfs for the type I-III and type II seesaw ensembles are shown in figure \ref{FigPDFN2den}.  The symmetry pattern of the pdfs under the exchange $\hat{m}_{1}\leftrightarrow\hat{m}_{2}$ is easily seen from figure \ref{FigPDFN2den}.

In fact, the plots are constructed in a way that takes advantage of this symmetry to provide a meaningful comparison between analytical and numerical results in both cases.  Indeed, by showing only half the data for the analytical $(\hat{m}_{1}<\hat{m}_{2})$ and numerical $(\hat{m}_{1}>\hat{m}_{2})$ results in each plot, it can be seen that the agreement between the two is once again very good.  Moreover, by comparing the distances between modes in each plot, it is possible to determine which ensemble has a stronger repulsion between the singular values.  It is found that the modes are $1.7$ times farther apart in the type I-III seesaw ensemble, meaning that a stronger repulsion is observed between the singular values for this ensemble.  Unfortunately, the origin of this interesting feature is hard to trace back without a completely-integrated analytical expression for $P_\nu^\text{I-III}(\hat{m}_1,\hat{m}_2)$.  A possible explanation as to why the singular values are closer together in the type II seesaw ensemble is suggested in expression \eqref{EqnPDFN2}.  Indeed, one can see that the Vandermonde-like contribution $|\hat{m}_1^2-\hat{m}_2^2|$ in $P_\nu^\text{I-III}(\hat{m}_1,\hat{m}_2)$ (responsible for the spreading of the singular values with reference to the symmetry axis), is strongly suppressed by a product of masses to the fifth power, which tends to spread the singular values in a narrow band along the two axes.  Eventually, in order to get a better understanding of the type I-III seesaw ensemble, the density plot could be used to guess an analytical form for $P_\nu^\text{I-III}(\hat{m}_1,\hat{m}_2)$ by fitting some appropriate functions of the singular values with free parameters to be determined.  Such density plots are not really conceivable for the case $N=3$ as they would require heavy numerical computation and would be rather hard to illustrate properly.  However, the same reasoning and conclusions apply to this case as well.

\subsection{The Case \texorpdfstring{$N=3$}{N=3}: SM Neutrino Physics}\label{SsN3}

With the tools and insights developed in the previous sections, it is now possible to fully analyze the more interesting case of a seesaw-extended SM.
\begin{table}[!t]
\centering
\resizebox{8cm}{!}{%
\begin{tabular}{|c||c|c|}
\hline
 & Normal Hierarchy & Inverted Hierarchy\\\hline
$\theta_{12}({}^\circ)$ & $33.48_{-0.75}^{+0.78}$ & $33.48_{-0.75}^{+0.78}$\\
$\theta_{23}({}^\circ)$ & $42.3_{-1.6}^{+3.0}$ & $49.5_{-2.2}^{+1.5}$\\
$\theta_{13}({}^\circ)$ & $8.50_{-0.21}^{+0.20}$ & $8.51_{-0.21}^{+0.20}$\\
$\delta({}^\circ)$ & $306_{-70}^{+39}$ & $254_{-62}^{+63}$\\
$\Delta m_{21}^2(10^{-5}\,\text{eV}^2)$ & $7.50_{-0.17}^{+0.19}$ & $7.50_{-0.17}^{+0.19}$\\
$\Delta m_{3\ell}^2(10^{-3}\,\text{eV}^2)$ & $2.457_{-0.047}^{+0.047}$ & $-2.449_{-0.047}^{+0.048}$\\
$m_1(\text{eV})$ & $<4.5$ & $<4.5$\\
\hline
\end{tabular}
}
\caption{Best-fit values for the SM neutrino physics parameters for the normal and inverted hierarchies.  The intervals correspond to $\pm1\sigma$.  In the case of $\Delta m_{3\ell}^2$, $\ell=1$ for the normal hierarchy and $\ell=2$ for the inverted hierarchy.}
\label{TabNeutrinos}
\end{table}

First, recent experimental values of the physical parameters in the neutrino sector are summarized in table \ref{TabNeutrinos}.  The mixing angles, the CP-violating Dirac phase and the squared mass differences $\Delta m_{ij}^2=m_i^2-m_j^2$ are extracted from \cite{Gonzalez-Garcia:2014bfa}.\footnote{While numerically computing the results of this paper, an updated fit of the neutrino sector experimental values was published in \cite{Esteban:2016qun}.  Since the experimental values did not change much, the analysis presented here will not change significantly.}  The upper bound on the mostly-electronic neutrino $m_1$ is the $1\sigma$ upper bound of \cite{Pagliaroli:2010ik}.  This upper bound on $m_1$ comes from the study of supernova.  It is conservative and quite model-independent.  Indeed, it is the weakest bound when compared to the cosmological bound on the sum of the neutrino masses which is somewhat model-dependent or the neutrinoless double $\beta$-decay bound and the direct neutrino mass bound which are intertwined with some mixing matrix parameters \cite{Drexlin:2013lha}.  These values will be used in the probability test at the end of this section.

For neutrino physics, the most convenient parametrization for the unitary group $U(3)$ is of course the PMNS mixing matrix \cite{Olive:2016xmw} for the light neutrino $U_\nu$,
\eqna{
U_\nu&=\left(\begin{array}{ccc}1&0&0\\0&\cos(\theta_{23})&\sin(\theta_{23})\\0&-\sin(\theta_{23})&\cos(\theta_{23})\end{array}\right)\left(\begin{array}{ccc}\cos(\theta_{13})&0&\sin(\theta_{13})e^{-i\delta}\\0&1&0\\-\sin(\theta_{13})e^{i\delta}&0&\cos(\theta_{13})\end{array}\right)\\
&\phantom{=}\hspace{0.5cm}\times\left(\begin{array}{ccc}\cos(\theta_{12})&\sin(\theta_{12})&0\\-\sin(\theta_{12})&\cos(\theta_{12})&0\\0&0&1\end{array}\right)\left(\begin{array}{ccc}1&0&0\\0&e^{i\alpha_{21}/2}&0\\0&0&e^{i\alpha_{31}/2}\end{array}\right).
}[EqnPMNS]
The group variable pdf \eqref{EqnPDF} is given by
\eqn{P_\nu(\theta_{12},\theta_{13},\theta_{23},\delta,\alpha_{21},\alpha_{31})=\frac{1}{2\pi^3}\sin(2\theta_{12})\sin(\theta_{13})[\cos(\theta_{13})]^3\sin(2\theta_{23}),}[EqnPDFN3]
which is the same as the normalized Haar measure \eqref{EqnHaar} obtained from the parametrization \eqref{EqnU}.  Hence the complex seesaw ensemble prefers the mixing angles $\theta_{12}$ and $\theta_{23}$ around $\pi/4$ and the mixing angle $\theta_{13}$ around $\pi/6$.  The pdfs for the CP-violating Dirac phase $\delta$ and the two CP-violating Majorana phases $\alpha_{21}$ and $\alpha_{32}$ are flat.  Therefore, any value for the CP-violating phases is equally probable in the complex seesaw ensemble.  It is important to note that the unphysical phases are not explicitly included in the PMNS parametrization \eqref{EqnPMNS}.
\begin{figure}[!t]
\centering
\resizebox{15cm}{!}{
\includegraphics{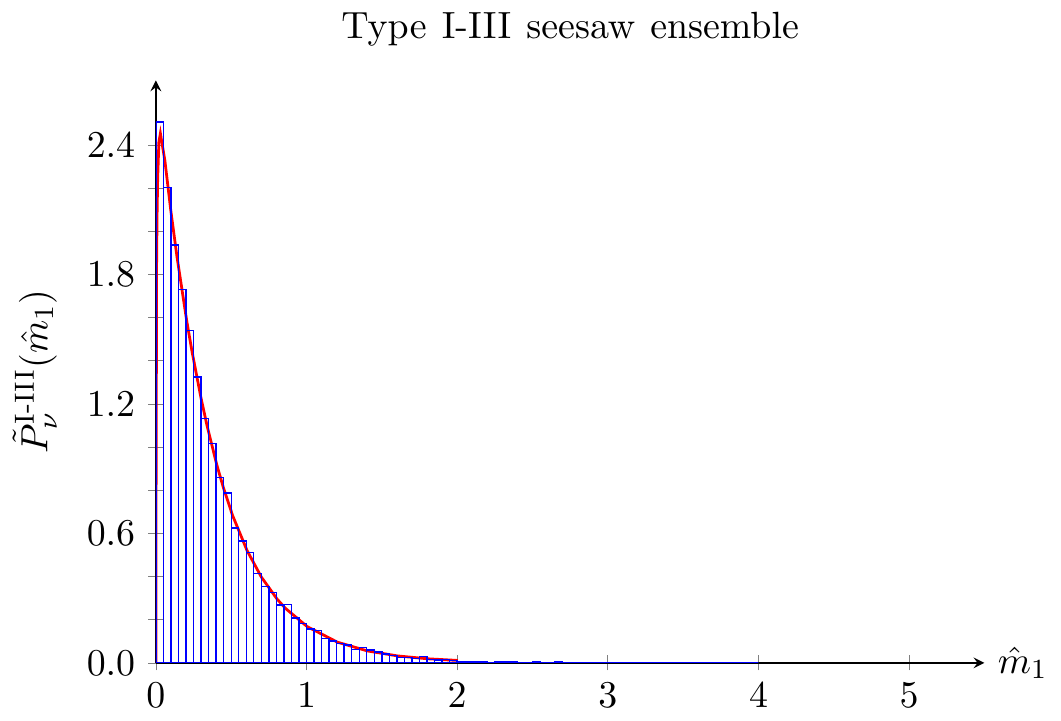}
\hspace{2cm}
\includegraphics{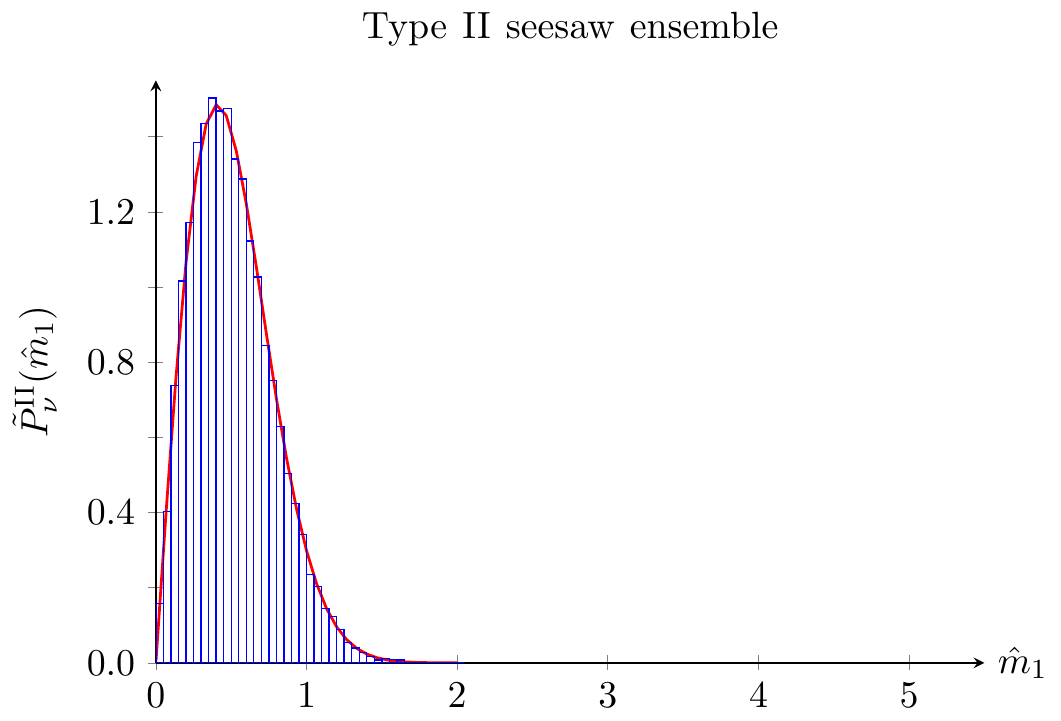}
}
\resizebox{15cm}{!}{
\includegraphics{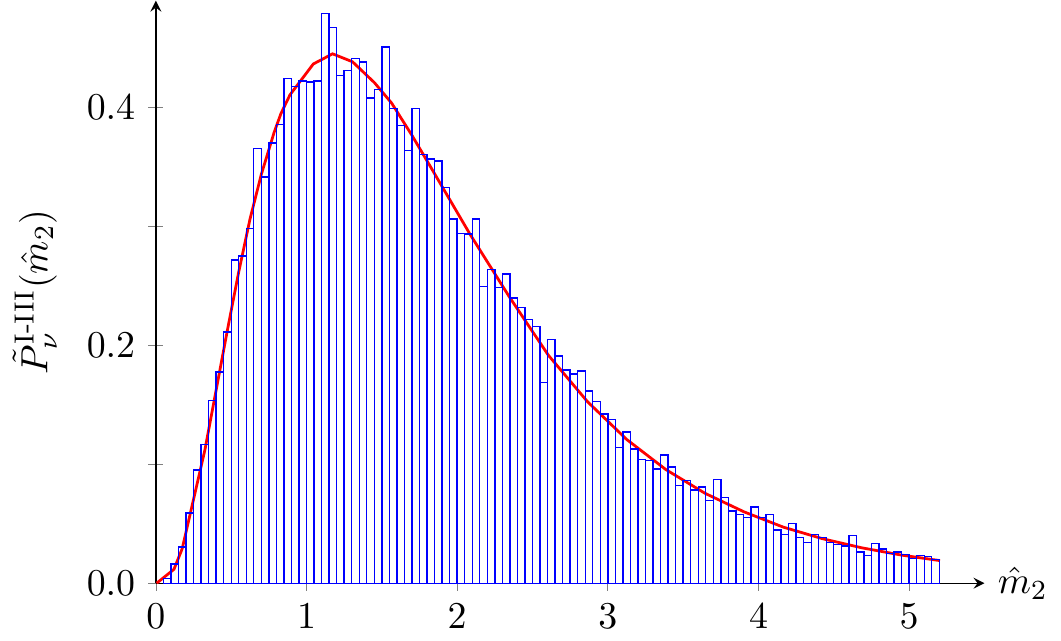}
\hspace{2cm}
\includegraphics{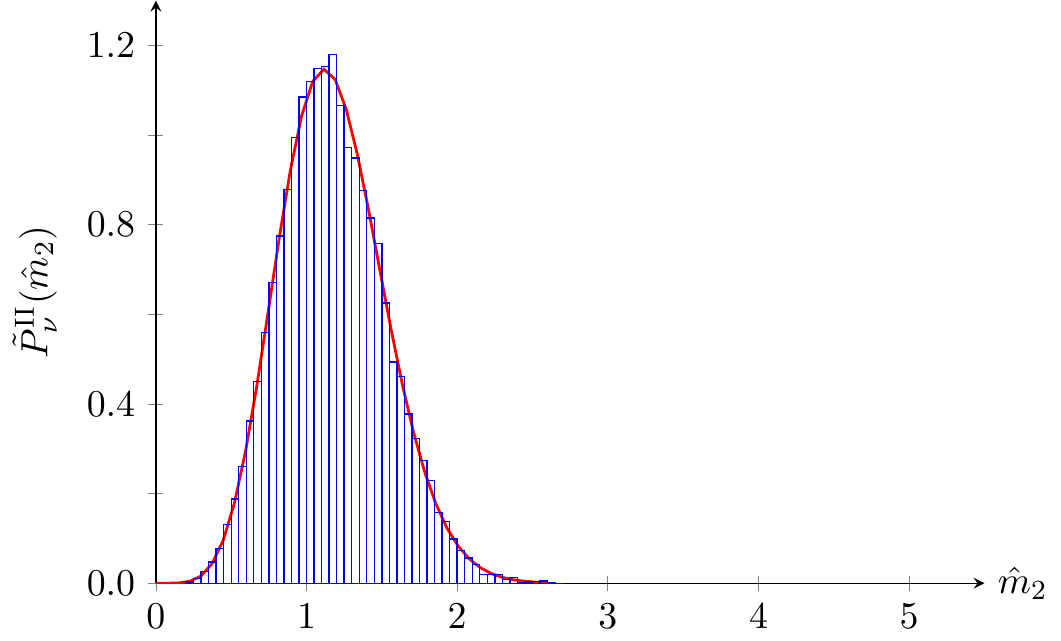}
}
\resizebox{15cm}{!}{
\includegraphics{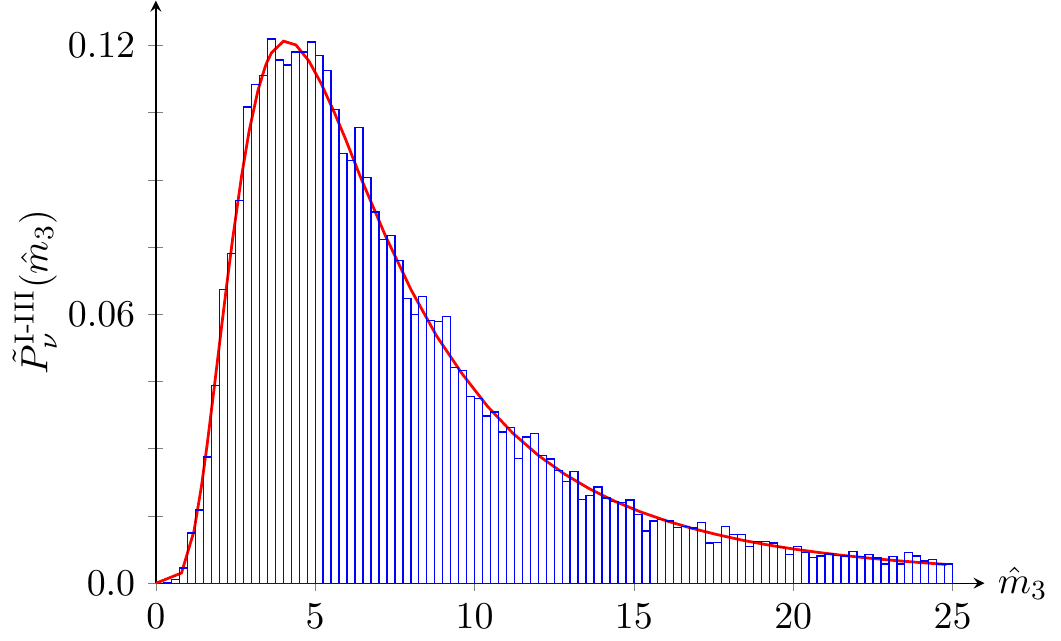}
\hspace{2cm}
\includegraphics{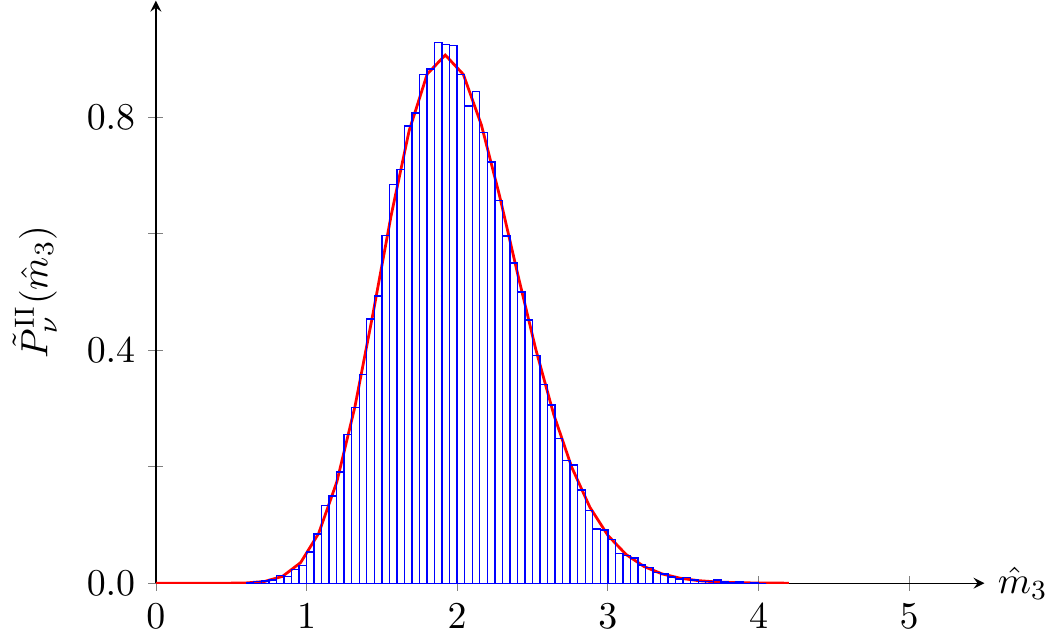}
}
\caption{Probability density functions for the singular values (masses) of the complex seesaw ensembles with $N=3$.  The red curve corresponds to the analytic result while the histogram corresponds to numerical results (with $2.5\times10^4$ dimensionless light neutrino mass matrices generated).  The left and right columns show the pdfs for the smallest, median and largest singular values for the type I-III and the type II seesaw mechanisms respectively.  The singular values are ordered such that $0\leq\hat{m}_1\leq\hat{m}_2\leq\hat{m}_3$ and an extra factor of $3!$ is introduced to correct the singular value pdfs.}
\label{FigPDFN3}
\end{figure}

The full expression for the type I-III singular value pdf is quite long (the explicit expression fills up a few pages) and not enlightening.  The form \eqref{EqnPDF} with the parametrization \eqref{EqnU} and $N=3$ is sufficient for both type I-III and type II.  As discussed in section \ref{SsConsequences}, these pdfs are invariant under permutations of the singular values $\hat{m}_1$, $\hat{m}_2$ and $\hat{m}_3$.  Consequently, the hierarchy of the neutrino mass spectrum of the extended SM cannot be predicted under the anarchy hypothesis.  The only claim that can be made is that all hierarchy scenarios (that is to say, every $3!$ permutations of the three singular values) are equiprobable for a given mass splitting (again, the term mass splitting is understood to be a particular ordering of the two quantities $\hat{m}^{2}_{\text{med}}-\hat{m}^{2}_{\text{min}}$ and $\hat{m}^{2}_{\text{max}}-\hat{m}^{2}_{\text{med}}$).  To further study the mass splitting and the resulting marginal pdfs, it is convenient to introduce a particular singular value ordering.  From now on, the ordering is chosen to be $0\leq\hat{m}_1\leq\hat{m}_2\leq\hat{m}_3$ (such that $\hat{m}_1\equiv\hat{m}_\text{min}$, $\hat{m}_2\equiv\hat{m}_\text{med}$ and $\hat{m}_3\equiv\hat{m}_\text{max}$) although it must remain clear that those are not straightforwardly related to the experimental neutrino masses (once again, the choice is completely arbitrary).  Thus, one cannot single out any of the two hierarchy scenarios selected by experimental data.  Nevertheless, by studying the marginal singular value pdfs of the type I-III and type II seesaw ensembles, some important and interesting results on the regime of low-energy neutrino physics can be obtained.

\begin{table}[!t]
\centering
\resizebox{8cm}{!}{%
\begin{tabular}{|c||c|c|c|}
\hline
Marginal pdfs & Mean & Median & Mode \\\hline
$\tilde{P}_\nu^\text{I-III}(\hat{m}_1)$ & $0.36$ & $0.26$ & $0.03$\\
$\tilde{P}_\nu^\text{I-III}(\hat{m}_2)$ & $1.93$ & $1.65$ & $1.17$\\
$\tilde{P}_\nu^\text{I-III}(\hat{m}_3)$ & $9.65$ & $6.55$ & $4.09$\\\hline
$\tilde{P}_\nu^\text{II}(\hat{m}_1)$ & $\frac{1}{2}\sqrt{\frac{\pi}{3}}$ & $0.48$ & $\frac{1}{\sqrt{6}}$\\
$\tilde{P}_\nu^\text{II}(\hat{m}_2)$ & $\frac{15}{16}\sqrt{\frac{\pi}{2}}$ & $1.15$ & $\frac{\sqrt{5}}{2}$\\
$\tilde{P}_\nu^\text{II}(\hat{m}_3)$ & $\frac{\sqrt{\pi }}{96} \left(72+45 \sqrt{2}-16 \sqrt{3}\right)$ & $1.96$ & $1.91$\\
\hline
\end{tabular}
}
\caption{Location parameters for the marginal singular value pdfs of figure \ref{FigPDFN3}.}
\label{TabparadistN3}
\end{table}
First, the marginal singular value pdfs can be obtained by computing the following integrals,
\eqna{
\tilde{P}_\nu^\varSigma(\hat{m}_1)&= 3!\int_{\hat{m}_1}^\infty d\hat{m}_2\int_{\hat{m}_2}^\infty d\hat{m}_3P_\nu^\varSigma(\hat{m}_1,\hat{m}_2,\hat{m}_3),\\
\tilde{P}_\nu^\varSigma(\hat{m}_2)&=3!\int_{\hat{m}_2}^\infty d\hat{m}_3\int_{0}^{\hat{m}_2} d\hat{m}_1P_\nu^\varSigma(\hat{m}_1,\hat{m}_2,\hat{m}_3),\\
\tilde{P}_\nu^\varSigma(\hat{m}_3)&=3!\int_{0}^{\hat{m}_3} d\hat{m}_2\int_{0}^{\hat{m}_2} d\hat{m}_1P_\nu^\varSigma(\hat{m}_1,\hat{m}_2,\hat{m}_3),
}
for both ensembles (\textit{i.e.} $\varSigma=\text{I-III}$ or $\text{II}$).  The marginal singular value pdfs are shown in figure~\ref{FigPDFN3}.  A first observation that comes to mind when looking at figure \ref{FigPDFN3} is the diversity of the mass spectrum obtained with the type I-III seesaw ensemble compared to the type II seesaw ensemble.  This can be traced back to the fact that the pdfs $\tilde{P}_\nu^\text{I-III}(\hat{m}_{i})$ are much more complex than the simple Gaussian-like pdfs $\tilde{P}_\nu^\text{II}(\hat{m}_{i})$ that arise in the type II seesaw ensemble.  For example, there is no internal angular dependence associated to extra variables that needs to be integrated out in the type II pdf.  A second observation worth mentioning is the remarkable agreement between analytical and numerical results.  For reasons previously mentioned, analytical results derived from the type I-III seesaw ensemble are much more challenging to get than those of the type II seesaw ensemble.  Following heavy numerical computation based on adaptive Monte Carlo integration, the $11$-dimensional integrals resulting from the marginalization procedure can be obtained for given values of $\hat{m}_1$, $\hat{m}_2$ or $\hat{m}_3$.  The red curves produced in such a way are in very good agreement with their corresponding histograms (generated from a sample of light neutrino mass matrices), which can be viewed as a validation of the Monte Carlo integration method (estimated to be accurate to at least three significant figures) used in this case or a numerical check of the singular value pdfs obtained in \cite{Fortin:2016zyf}.

\begin{figure}[!t]
\centering
\resizebox{15cm}{!}{
\includegraphics{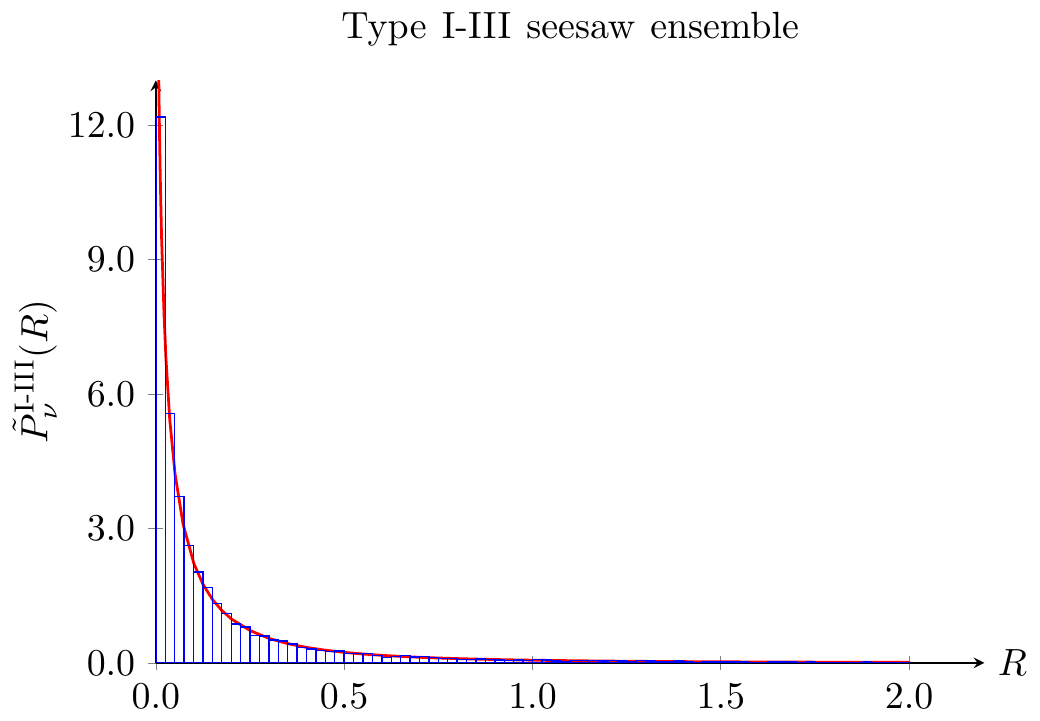}
\hspace{2cm}
\includegraphics{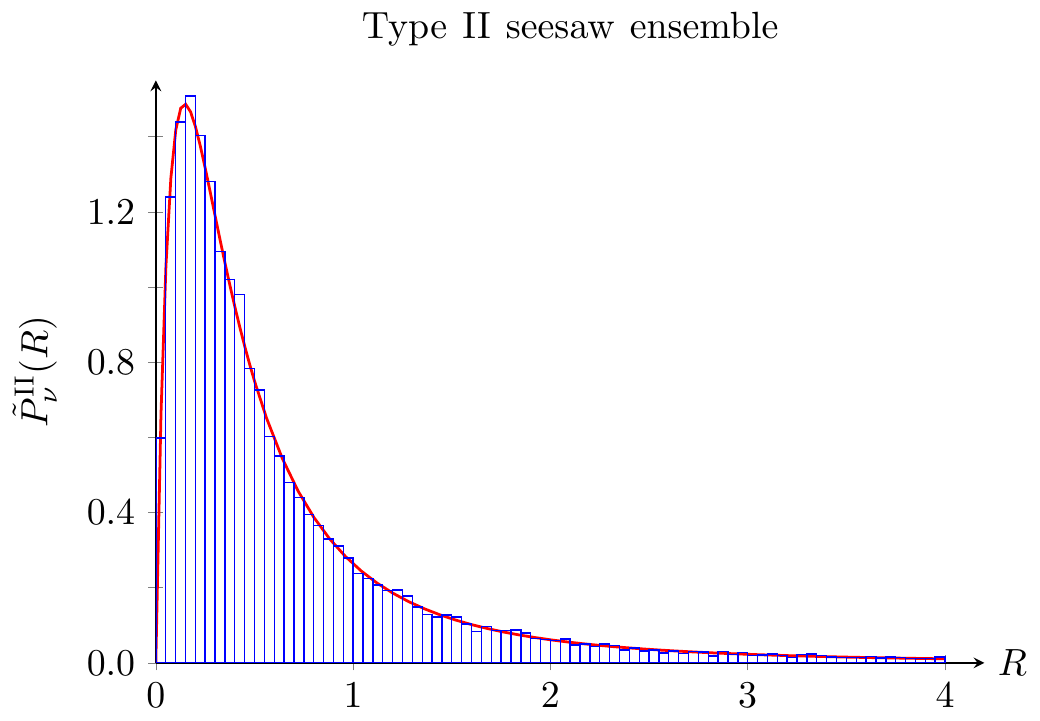}
}
\caption{Probability density functions for the ratios of the complex seesaw ensembles with $N=3$.  The red curve corresponds to the analytic result while the histogram corresponds to numerical results (with $2.5\times10^4$ dimensionless light neutrino mass matrices generated).  The left and right plots show the pdfs for the ratios of the type I-III and the type II seesaw mechanisms respectively.  The singular values are ordered such that $0\leq\hat{m}_1\leq\hat{m}_2\leq\hat{m}_3$ and an extra factor of $3!$ is introduced to correct the singular value pdfs.}
\label{FigPDFr}
\end{figure}
Once the numerical integration method is carefully tested, the next step is to compute the relevant statistical parameters.  The results are presented in table \ref{TabparadistN3}.  Following the same approach as in the $N=2$ case, it is found that the average singular values are once again spread over a much wider range in the type I-III seesaw ensemble.  Moreover, when compared to the $N=2$ case, it can be seen that this range expands significantly more in the type I-III seesaw ensemble as $N$ increases.  Next, comparing these values with their respective medians, one can conclude that the pdfs are much more symmetric in the type II seesaw ensemble, as can be expected when looking at figure \ref{FigPDFN3}.

Even though determining the hierarchy of the mass spectrum is out of reach in the context of the seesaw ensembles, the previous results can still be used to help identify which of the two possible mass splittings (according to our previous ordering, the mass splittings can be written as $\Delta \hat{m}_{21}^2=\hat{m}^{2}_{\text{med}}-\hat{m}^{2}_{\text{min}}$ and $\Delta \hat{m}_{32}^2=\hat{m}^{2}_{\text{max}}-\hat{m}^{2}_{\text{med}}$ so that the two possibility are either $\Delta\hat{m}_{21}^2<\Delta\hat{m}_{32}^2$ or $\Delta\hat{m}_{21}^2>\Delta\hat{m}_{32}^2$) is more likely to occur under the anarchy hypothesis.

\begin{figure}[!t]
\centering
\resizebox{15cm}{!}{
\includegraphics{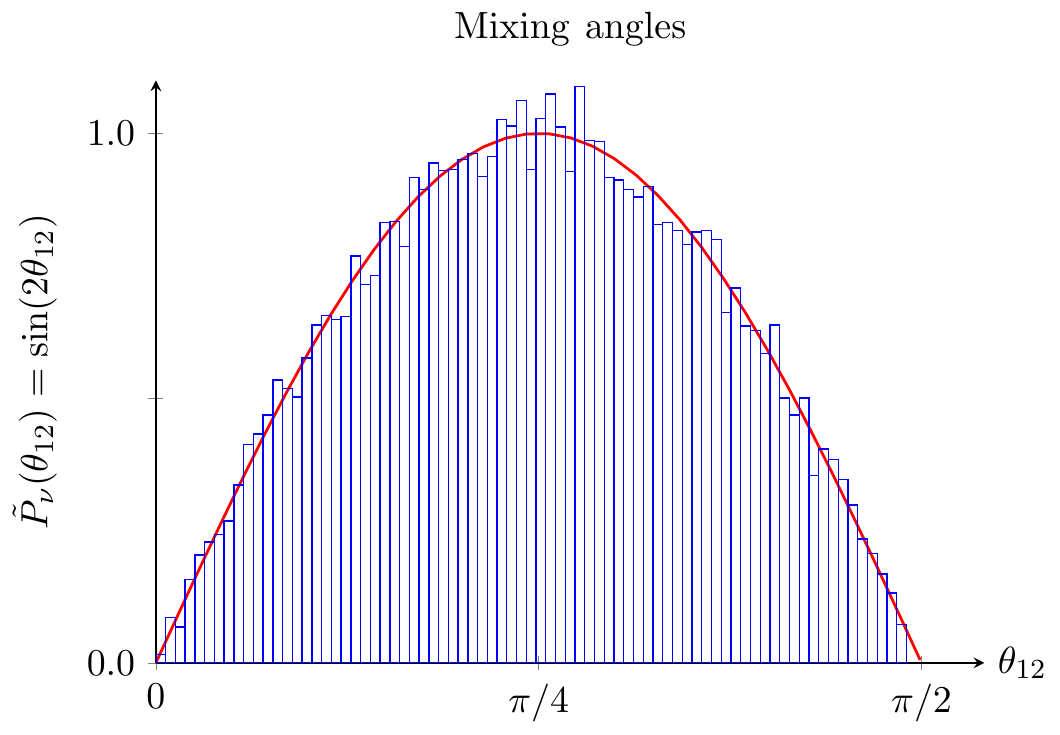}
\hspace{2cm}
\includegraphics{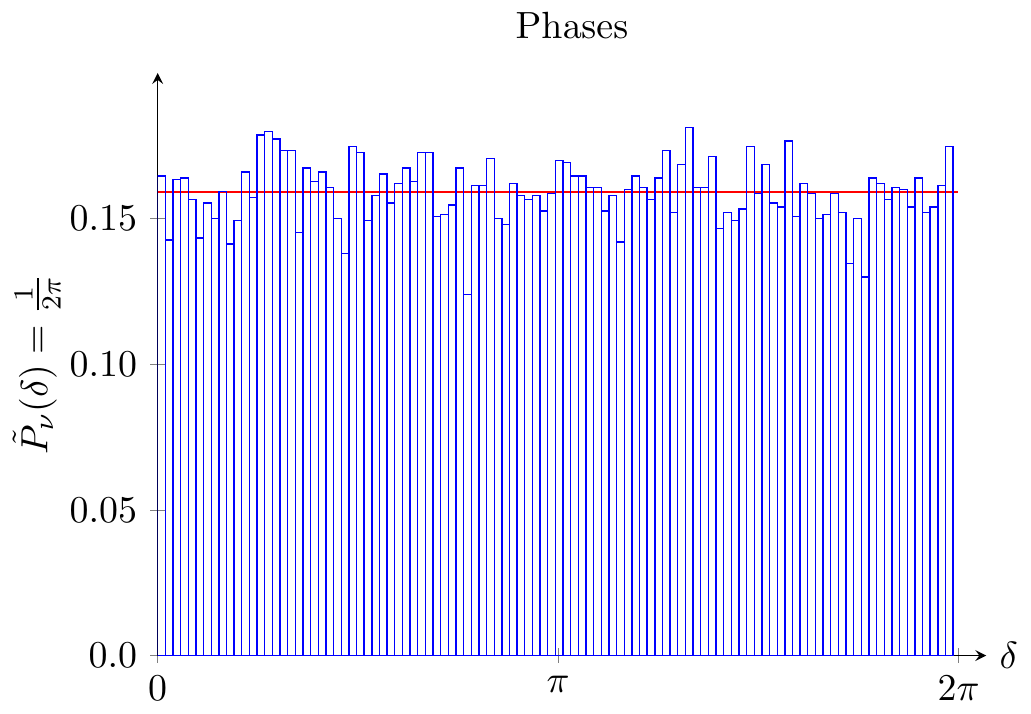}
}
\resizebox{15cm}{!}{
\includegraphics{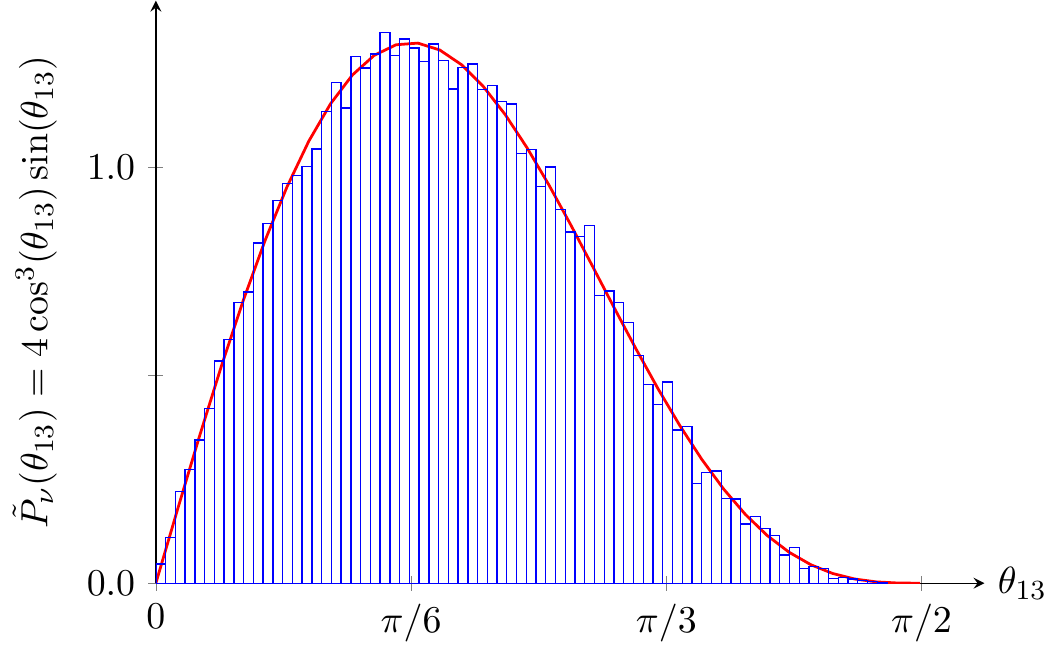}
\hspace{2cm}
\includegraphics{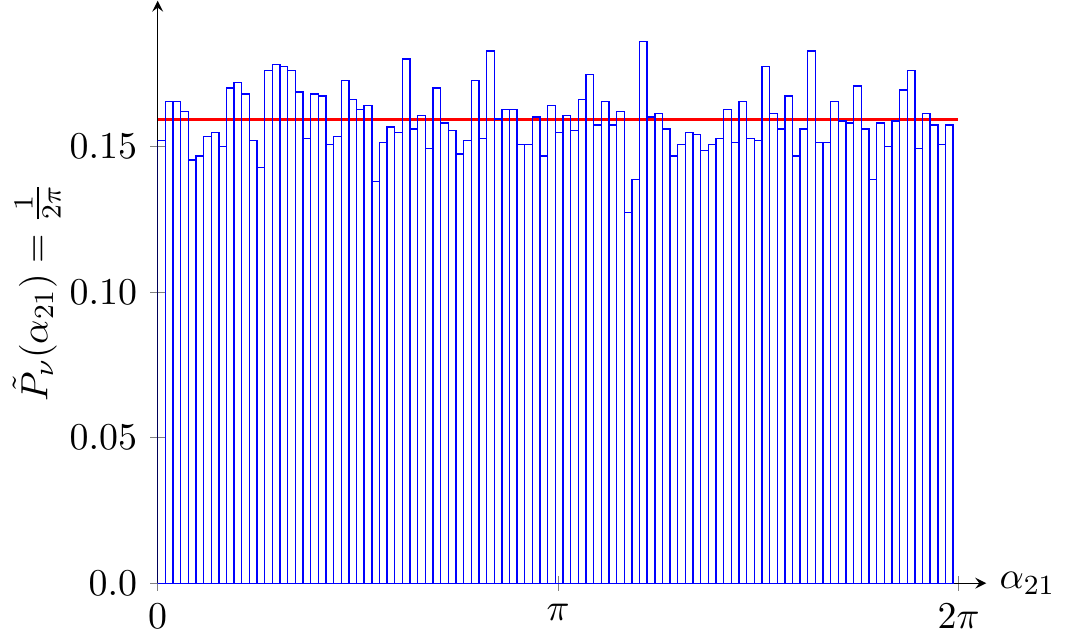}
}
\resizebox{15cm}{!}{
\includegraphics{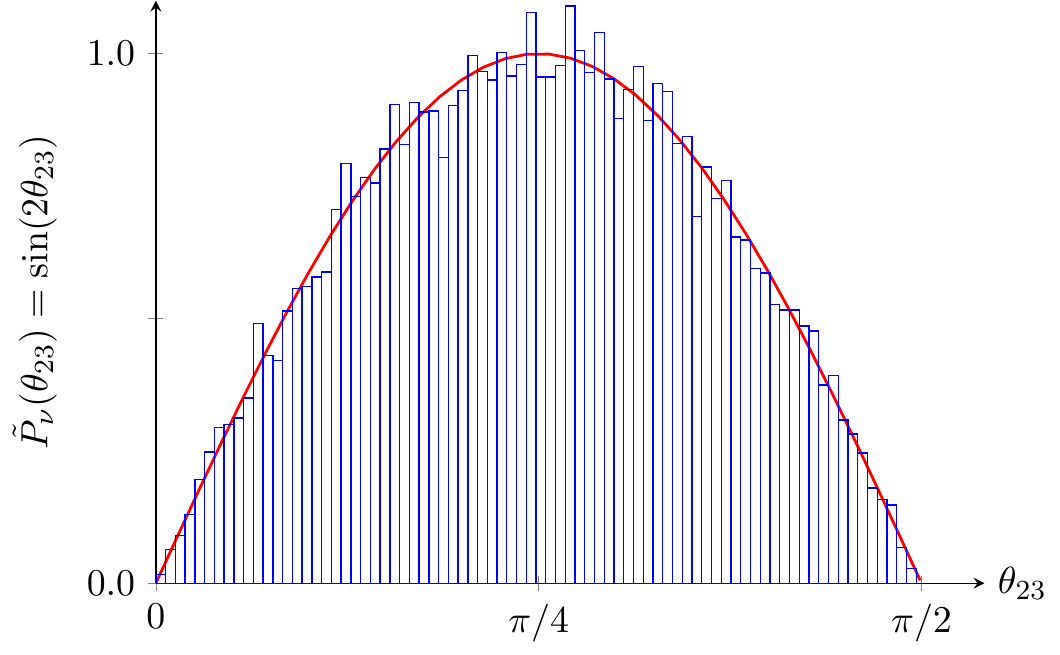}
\hspace{2cm}
\includegraphics{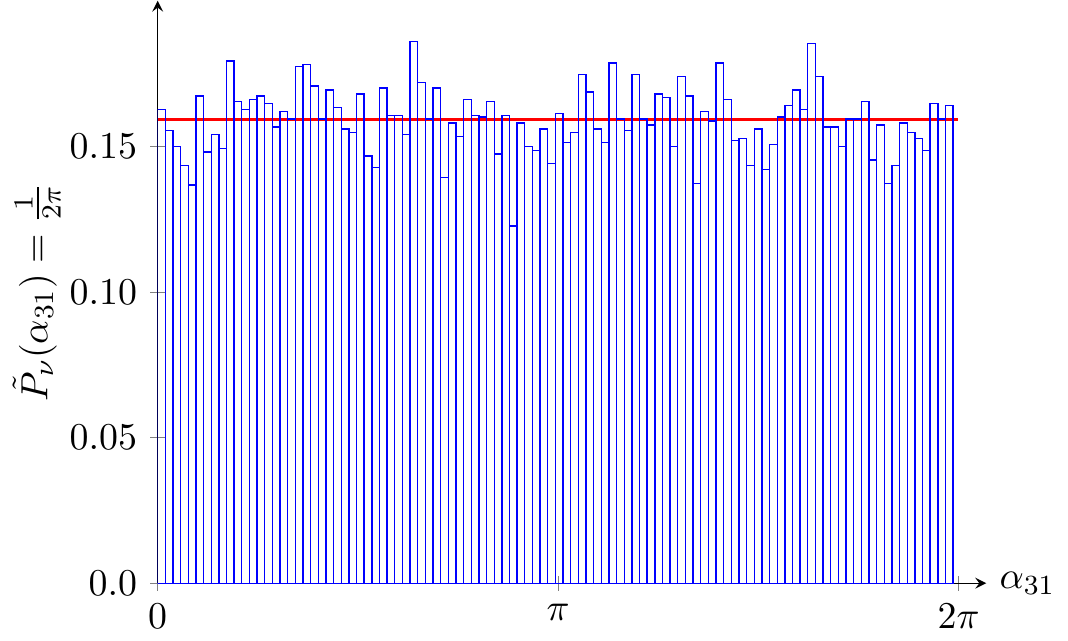}
}
\caption{Probability density functions for the mixing angles and phases of the complex seesaw ensembles with $N=3$.  The red curve corresponds to the analytic result while the histogram corresponds to numerical results (with $2.5\times10^4$ dimensionless light neutrino mass matrices generated).  The left and right columns show the mixing angles and the remaining flat phases respectively.  Since these distributions depend only on the Haar measure of the corresponding Lie group [$U(3)$ in this case], there is no distinction between type I-III and type II seesaw mechanisms.}
\label{FigPDFang}
\end{figure}
By studying the distribution of the ratio $R$,
\eqn{R=\frac{\Delta \hat{m}_{21}^2}{\Delta \hat{m}_{32}^2}=\frac{\hat{m}_{2}^{2}-\hat{m}_{1}^{2}}{\hat{m}_{3}^{2}-\hat{m}_{2}^{2}},}
which leads to the marginal pdf $\tilde{P}_\nu^\varSigma(R)$ for both ensembles, it becomes clear that the pdf resulting from type I-III seesaw ensemble is more likely to reproduce the experimental value of $R_{\text{exp}}\simeq0.03$ for the normal hierarchy (also when compared to $R_{\text{exp}}\simeq32.65$ for the inverted hierarchy).  Moreover, by integrating these distributions over the range $0\leq R\leq1$ ($1\leq R\leq\infty$), one gets the probability that the mass splitting $\Delta\hat{m}_{21}^2<\Delta\hat{m}_{32}^2$ ($\Delta\hat{m}_{21}^2>\Delta\hat{m}_{32}^2$) is realized.  For the type I-III seesaw ensemble, the probability is $95.8\%$ ($4.2\%$) whereas for the type II seesaw ensemble the probability is $79.0\%$ ($21.0\%$).  These probabilities are supported by figure \ref{FigPDFr}.  One can thus conclude that the dominant trend for both ensembles is the realization of the mass splitting $\Delta \hat{m}_{21}^2<\Delta \hat{m}_{32}^2$ reminiscent of the normal hierarchy.  Moreover, looking at the very distinct behavior of the two pdfs and comparing the resulting probability, it is possible to state that the type I-III seesaw ensemble is better suited to generate this particular mass splitting.  In the context of the type I-III seesaw ensemble, this means that the mass differences are way more likely to coincide with the ones from normal hierarchy, yet the ordering of the masses is still unknown (once again, every $3!$ permutations are equally probable for this particular splitting).

Next, considering the group variable pdf for the mixing angles and phases \eqref{EqnPDFN3}, similar conclusions as with the case $N=2$ can be drawn.  Once again, the phases have flat distributions and two of the three mixing angles prefer near-maximal values as shown in figure \ref{FigPDFang}.  Here, the unphysical phases were not considered in the making of figure \ref{FigPDFang} as they were deemed not interesting for the present discussion.  The numerical data coming from a sample of light neutrino mass matrices is once again consistent with the marginal pdfs obtained from the Haar measure.  The only non-symmetric pdf for the mixing angles is the one associated to $\theta_{13}$.  Its mode, located at $\pi/6$, is in agreement with the results of section \ref{SsConsequences} obtained with the parametrization \eqref{EqnU} since the Haar measure is the same for the PMNS matrix \eqref{EqnPMNS}.

To further emphasize the differences between the two ensembles, the probability test is now used to determine how well the complex seesaw ensembles can generate the observed values of the extended SM physical parameters in the neutrino sector.  The results of the probability test will then be compared between the type I-III and type II seesaw ensembles to determine which ensemble is more likely to generate the observed values of physical parameters.  The experimental values are given in table \ref{TabNeutrinos}.

\begin{figure}[!t]
\centering
\resizebox{15cm}{!}{
\includegraphics{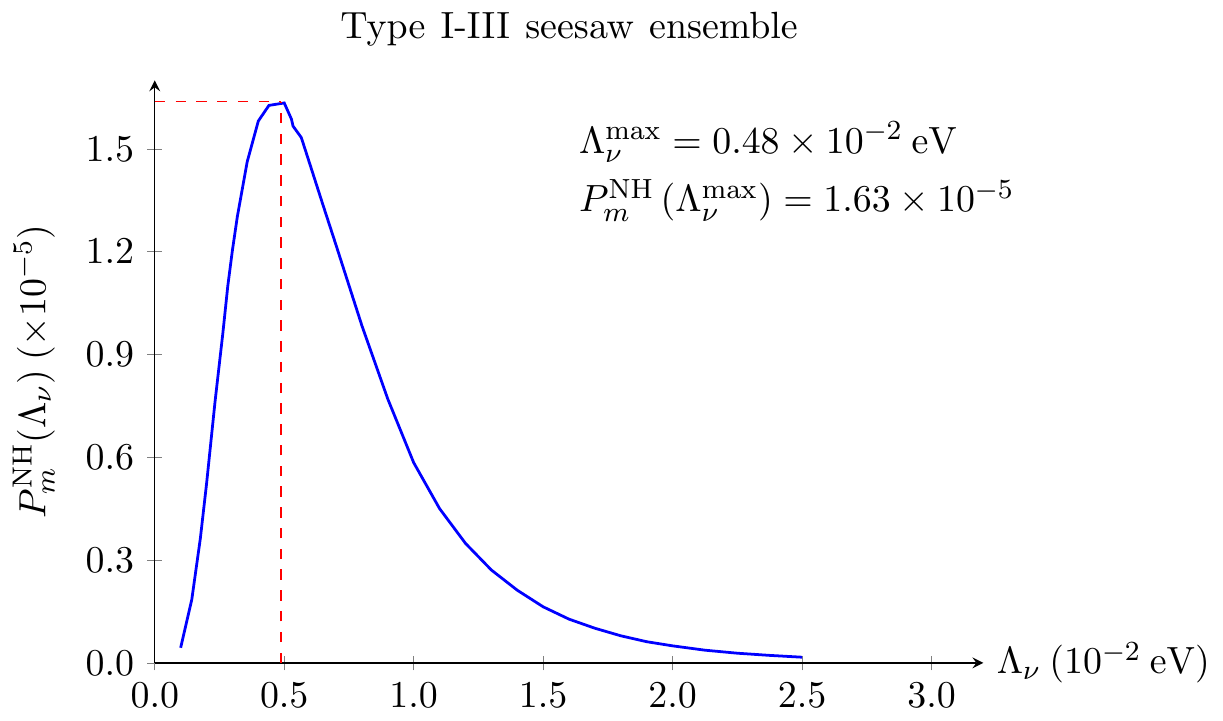}
\hspace{2cm}
\includegraphics{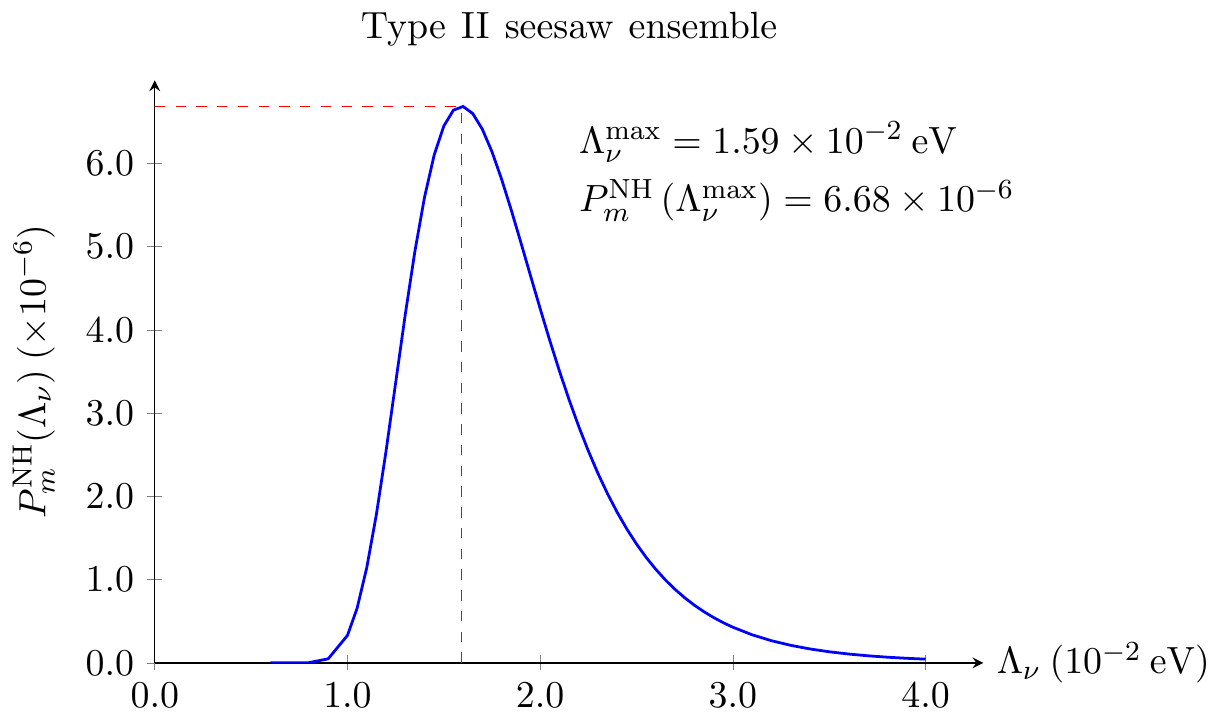}
}
\resizebox{15cm}{!}{
\includegraphics{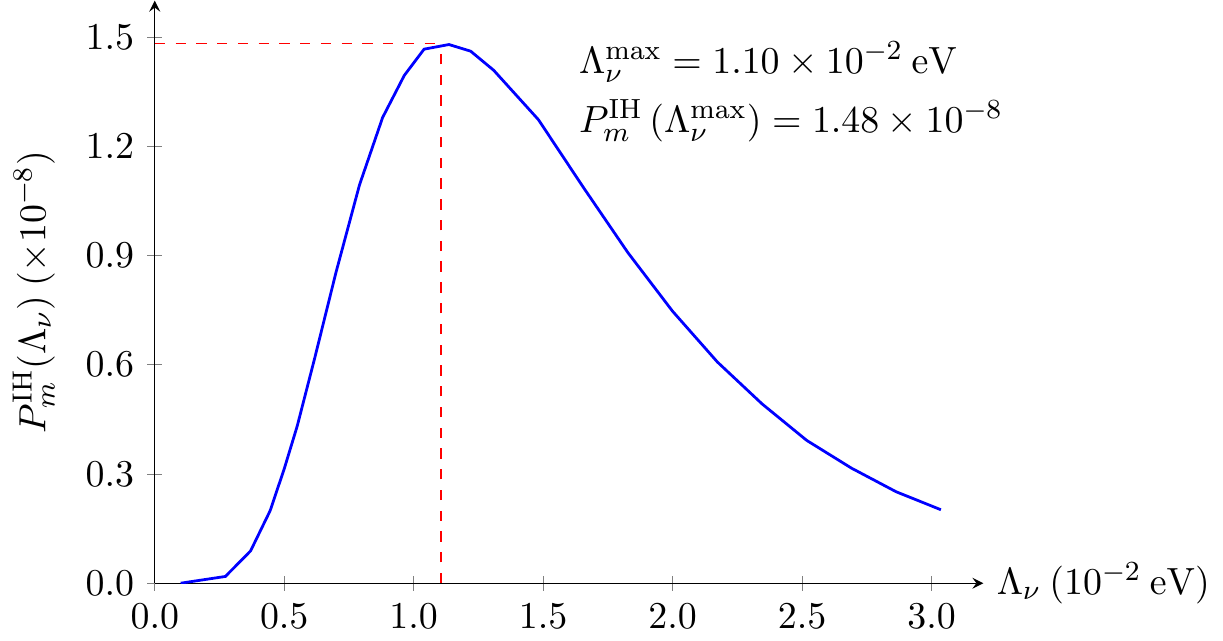}
\hspace{2cm}
\includegraphics{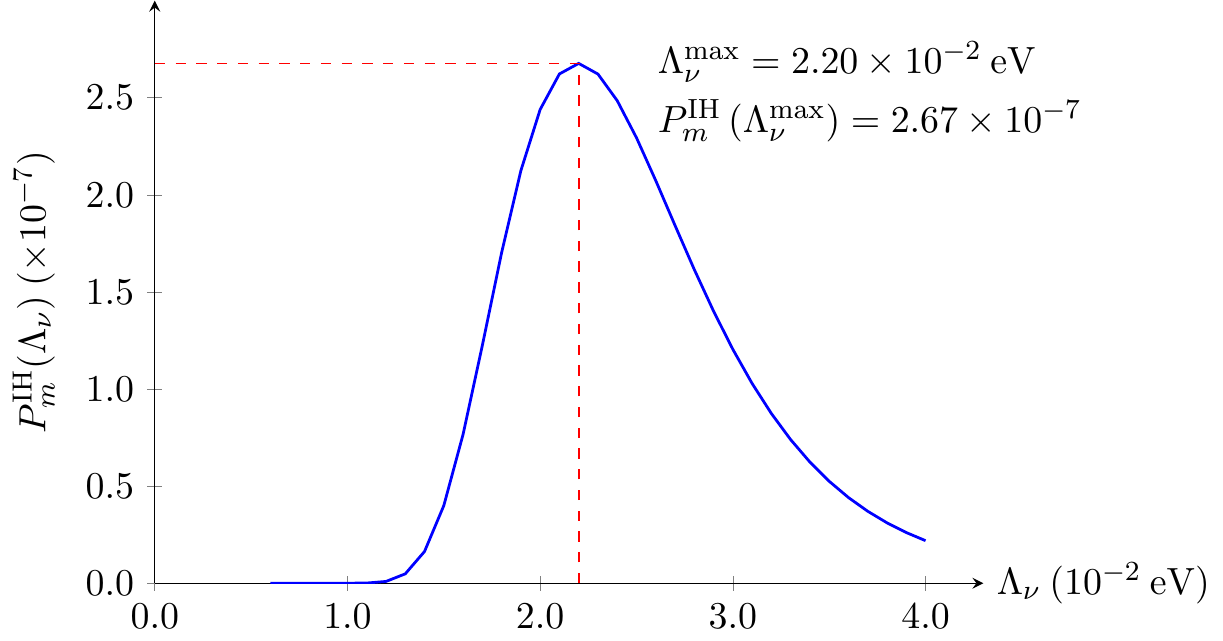}
}
\caption{Probability test for the singular values of the complex seesaw ensembles with $N=3$.  The left and right columns show the probability distribution as a function of $\Lambda_\nu$ (in the normal and inverted hierarchy scenarios) for the type I-III and the type II seesaw mechanisms respectively.  The singular values are in no particular order for this test.}
\label{FigPtest}
\end{figure}

With the observed values of table \ref{TabNeutrinos}, the probabilities (where $|\text{det}J|$ is the Jacobian of the appropriate hierarchy) are
\eqna{
P_m^\text{NH}(\Lambda_\nu)&=\int_{\frac{(7.50-0.17)\times10^{-5}\,\text{eV}^2}{2\Lambda_\nu^2}}^{\frac{(7.50+0.19)\times10^{-5}\,\text{eV}^2}{2\Lambda_\nu^2}}d\Delta\hat{m}_{21}^2\int_{\frac{(2.457-0.047)\times10^{-3}\,\text{eV}^2}{2\Lambda_\nu^2}}^{\frac{(2.457+0.047)\times10^{-3}\,\text{eV}^2}{2\Lambda_\nu^2}}d\Delta\hat{m}_{31}^2\\
&\phantom{=}\hspace{2cm}\times\int_0^{\frac{4.5\,\text{eV}}{\sqrt{2}\Lambda_\nu}}d\hat{m}_{1}|\text{det}J|P_\nu^\varSigma(\hat{m}_1,\hat{m}_2,\hat{m}_3),\\
P_m^\text{IH}(\Lambda_\nu)&=\int_{\frac{(7.50-0.17)\times10^{-5}\,\text{eV}^2}{2\Lambda_\nu^2}}^{\frac{(7.50+0.19)\times10^{-5}\,\text{eV}^2}{2\Lambda_\nu^2}}d\Delta\hat{m}_{21}^2\int_{\frac{(-2.449-0.047)\times10^{-3}\,\text{eV}^2}{2\Lambda_\nu^2}}^{\frac{(-2.449+0.048)\times10^{-3}\,\text{eV}^2}{2\Lambda_\nu^2}}d\Delta\hat{m}_{32}^2\\
&\phantom{=}\hspace{2cm}\times\int_0^{\frac{4.5\,\text{eV}}{\sqrt{2}\Lambda_\nu}}d\hat{m}_{1}|\text{det}J|P_\nu^\varSigma(\hat{m}_1,\hat{m}_2,\hat{m}_3),\\
P_U&=\int_{V_\text{exp}}d\theta_{12}d\theta_{13}d\theta_{23}d\delta d\alpha_{21}d\alpha_{31}P_\nu(\theta_{12},\theta_{13},\theta_{23},\delta,\alpha_{21},\alpha_{31}),
}[ptest]
where NH stands for normal hierarchy while IH stands for inverted hierarchy.  The first test is achieved by using only the singular value pdfs for both ensembles (see figure \ref{FigPtest}).  For the type I-III seesaw ensemble, the $12$-dimensional integrals over the experimental volume defined by the $1\sigma$ range are obtained using the same Monte Carlo algorithm.  At this point, it is necessary to stress that this test does not require any ordering of the singular values.  In fact, all permutations are accounted for in these integrals and there is thus no need for an extra factor of $3!$ to ensure the normalisation of the pdfs.  Since the only free parameter left in the equations is $\Lambda_\nu$, the idea is to plot the probability as a function of $\Lambda_\nu$ over a range were the curves reach a maximum.  This allows for a simple comparison of their maximum values (by taking appropriate ratios) to determine the likelihood of each ensemble to generate the observed values.  It is important to note that beside the fact that the probabilities obtained this way are invariant under a change of basis, the explicit values of the probabilities are not particularly meaningful.  In fact, they are bound to shrink further and further as the experimental values get more and more precise.  However, the ratios are considered to be relevant quantities since they are subject to only small fluctuations during this process (the order of magnitude should remain the same).  The results of the probability test are presented in figure \ref{FigPtest}.

First, from within the same ensemble, one can compare the probabilities obtained from the normal and inverted hierarchy scenarios.  The maximum probability values resulting from a scan over $\Lambda_{\nu}$ reveal that the mass splitting in \eqref{ptest} are $\sim1000$ times more likely to originate from the region defined by normal hierarchy (at $1\sigma$) than from the one defined by inverted hierarchy in the type I-III seesaw ensemble.  In other words, this means that the type I-III seesaw ensemble is way more likely to generate values for these physical parameters that are contained within the region allowed by the normal hierarchy data set (rather than the inverted hierarchy data set).  For the type II seesaw ensemble, the same tendency is observed but with a much smaller ratio between the maximum probability values.  Indeed, figure~\ref{FigPtest} shows that the mass splitting is $\sim25$ times more likely to originate from the region defined by normal hierarchy.  It is then possible to conclude that between the two regions scanned in the probability test, both ensembles naturally lead to preferred values for these physical parameters that lie in the region defined by normal hierarchy.  Moreover, this preference is strongly accentuated in the type I-III seesaw ensemble.

Second, one can make a comparison between the two ensembles based on the result of figure~\ref{FigPtest}.  Since it was shown that one region is actively preferred over the other, it becomes useful to compare the maximum probability values in the case of normal hierarchy for both ensembles.  This time, the conclusions are not as striking as in the previous case but one can state that the type I-III seesaw ensemble is roughly $2$ times better than type II for generating values of these parameters in this particular region.\footnote{It is interesting to note that the type II seesaw ensemble is approximately $18$ times more probable than the type I-III seesaw ensemble for the inverted hierarchy.}  The results obtained from this probability test are thus in agreement with what was found previously by comparing the pdfs of the ratios $R$ and consequently help quantify the underlying trends in both ensembles.

A final point of interest regarding this particular test concerns the energy scale $\Lambda_\nu$.  Again from figure \ref{FigPtest} one can see that choosing the integration region to be over the accepted experimental values (within the $1\sigma$ confidence level) naturally fixes the energy scale of the models.  Each scan shows that the maximum probability values are attained for values of $\Lambda_{\nu}$ which are of the same order of magnitude, namely $\Lambda_{\nu}\sim\mathcal{O}(10^{-2})\:\text{eV}$ for both ensembles.   Since $\Lambda_{\nu}$, which corresponds to the light neutrino mass scale, takes the general form $\Lambda_{\nu}=v^2/\Lambda_{\text{new}}$ for each type of seesaw mechanisms, with $v\simeq246\:\text{GeV}$ the usual Higgs vacuum expectation value, a quick estimate of the new energy scale $\Lambda_{\text{new}}$ associated to the particle content introduced in the extended SM with type I, type II or type III seesaw mechanism (right-handed neutrinos singlets, Higgs triplet and fermionic triplets respectively) can be made.  Indeed, using the previously-mentioned values, one gets $\Lambda_{\text{new}}\sim\mathcal{O}(10^{15})\:\text{GeV}$ for the new energy scale of the extended SM, which is very close to the energy scale of grand unified theory (GUT).  Naturally, $\Lambda_{\text{new}}$ is directly related to the masses of these newly-introduced particles.  However, in order to assess their corresponding mass scales, one needs to specify the order of magnitude of the coupling constants arising from each seesaw scenario.  The usual approach is to set the coupling constants to be of $\mathcal{O}\left(1\right)$ since there is no fundamental principle or symmetry pattern that require particularly small couplings.  This in turn suggests that the new particles introduced in the SM are quite heavy since $\Lambda_{\text{new}}$ becomes essentially their corresponding mass scale.  In fact, this result is typical of seesaw mechanisms and is often regarded as a prerequisite (when taking the naturalness argument into consideration to avoid seesaw-induced fine-tuning or hierarchy problems) for these mechanisms to give sensible predictions concerning the light neutrino masses.  It is however possible to lower the mass scales by simply postulating smaller coupling constants, somewhat disregarding the naturalness argument.  Overall, the results of the probability test are therefore consistent with high-energy phenomenology of the seesaw-extended SM.

\begin{table}[t]
\centering
\resizebox{8cm}{!}{%
\begin{tabular}{|c||c|c|}
\hline
 & Normal Hierarchy & Inverted Hierarchy\\\hline
$P_{U}$ & $2.435\times10^{-6}$ & $2.231\times10^{-6}$\\
$P_{\text{flat}}$ & $1.198\times10^{-6}$ & $1.105\times10^{-6}$\\
$P_{U}/P_{\text{flat}}$ & $2.031$ & $2.018$\\
\hline
\end{tabular}
}
\caption{Probability test for the mixing angles and phases of the complex seesaw ensembles with $N=3$ and a (trivial) normalized flat distribution.  The probability $P_{U}$ and $P_{\text{flat}}$ are obtained by integrating the normalized Haar measure and the flat distribution over the experimental volume $V_{\text{exp}}$ defined by the data at the $1\sigma$ confidence level (see table \ref{TabNeutrinos}).}
\label{Tabprobtest}
\end{table}
The second probability test, with results shown in table \ref{Tabprobtest}, concerns the mixing parameters of the neutrino sector, namely the mixing angles and phases of table~\ref{TabNeutrinos}.  In this case, the analysis is much simpler since there is no free parameter with which to scan a particular region and the pdf (the Haar measure) is also a lot less complicated.  Since both ensembles have the same pdf for the mixing angles and phases, a comparison between the two is not possible.  However, it is interesting to see how well these ensembles perform when compared to a trivial normalized flat distribution.  Here, the idea is simply to test whether there is any improvement when generating parameter values from the Haar measure obtained in the seesaw ensembles as opposed to a less interesting model where there would be no information or explicit dependence on the angular part in the pdf.  By comparing the probabilities that the generated values lie within $V_{\text{exp}}$ in both cases and for the two types of hierarchy, one gets the results of table~\ref{Tabprobtest}.  The first conclusion that can be drawn from these results is that, up to the level of accuracy acknowledged for this test (remember that this analysis remains sensible to the choice of integration volume to some extent), there can be no distinction between normal or inverted hierarchy.  Both regions are thus equally probable.  However, when comparing $P_{U}$ with $P_{\text{flat}}$, there is indeed improvement as the seesaw ensembles are essentially $\sim2$ times more likely to generate parameters within the allowed experimental region (for both hierarchy scenarios) than the flat distribution.

\subsection{Large N comparison}\label{SslargeN}

To follow up on our previous work \cite{Fortin:2016zyf} regarding the close resemblance of the type I-III singular value pdf at $N=1$ and the level density at large $N$, this section further investigates this connection by adding the comparison with the pdfs at $N=2$ and $N=3$ for the type I-III seesaw ensemble.  The starting point for an appropriate comparison of these quantities is the correlation function
\eqn{\rho_{\nu N}^{\text{I-III}}(x)=N\int P_{\nu}^{\text{I-III}}(x,\hat{m}_{2},\cdots,\hat{m}_{N})\prod_{2\leq i\leq N}d\hat{m}_{i},}
for $N=2$ and $N=3$ respectively.  Introducing a convenient rescaling of the variable $x\rightarrow\sqrt{N}\hat{m}_{\nu}$, the resulting correlation functions
\eqn{\hat{\rho}_{\nu 2}^{\text{I-III}}(\hat{m}_{\nu})=\sqrt{2}\int_{0}^{\infty} d\hat{m}_2 P_{\nu}^{\text{I-III}}(\hat{m}_{\nu},\hat{m}_{2}),\quad\quad\hat{\rho}_{\nu 3}^{\text{I-III}}(\hat{m}_{\nu})=\sqrt{3}\int_{0}^{\infty} d\hat{m}_2\int_{0}^{\infty} d\hat{m}_3 P_{\nu}^{\text{I-III}}(\hat{m}_{\nu},\hat{m}_{2},\hat{m}_{3}),}
with $\hat{\rho}(\hat{x})=\rho(x)/\sqrt{N}$, can be compared directly to the large $N$ histogram ($N=60$).  The resulting $11$-dimensional integral for $N=3$ is carried out using the previously-mentioned Monte Carlo algorithm.
\begin{figure}[!t]
\centering
\resizebox{15cm}{!}{
\includegraphics{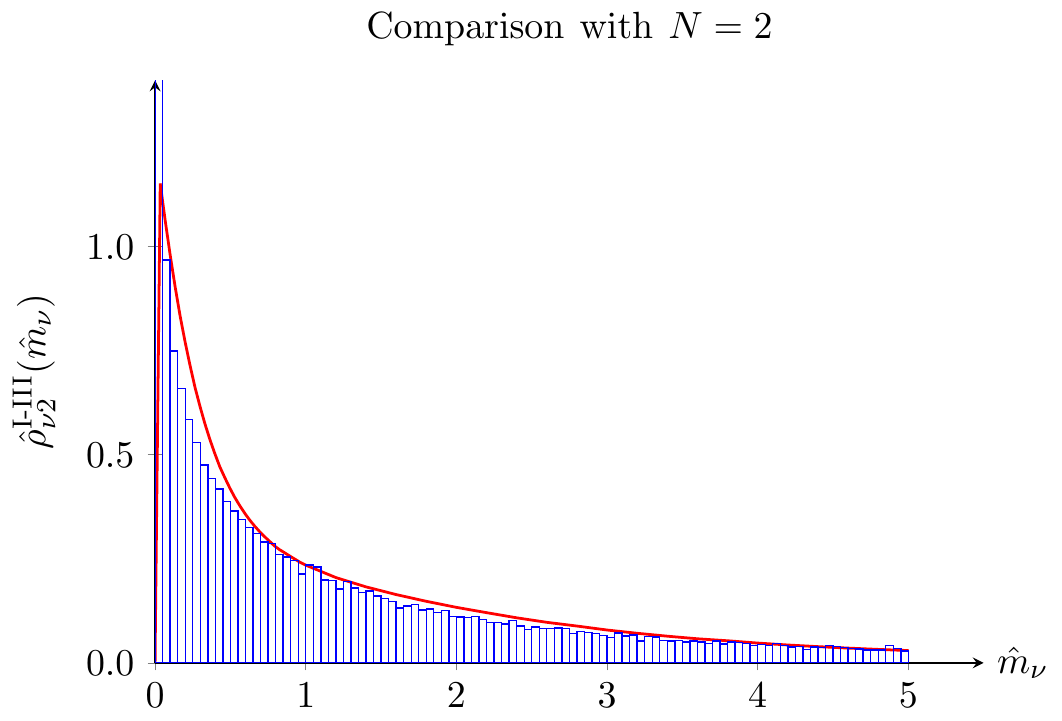}
\hspace{2cm}
\includegraphics{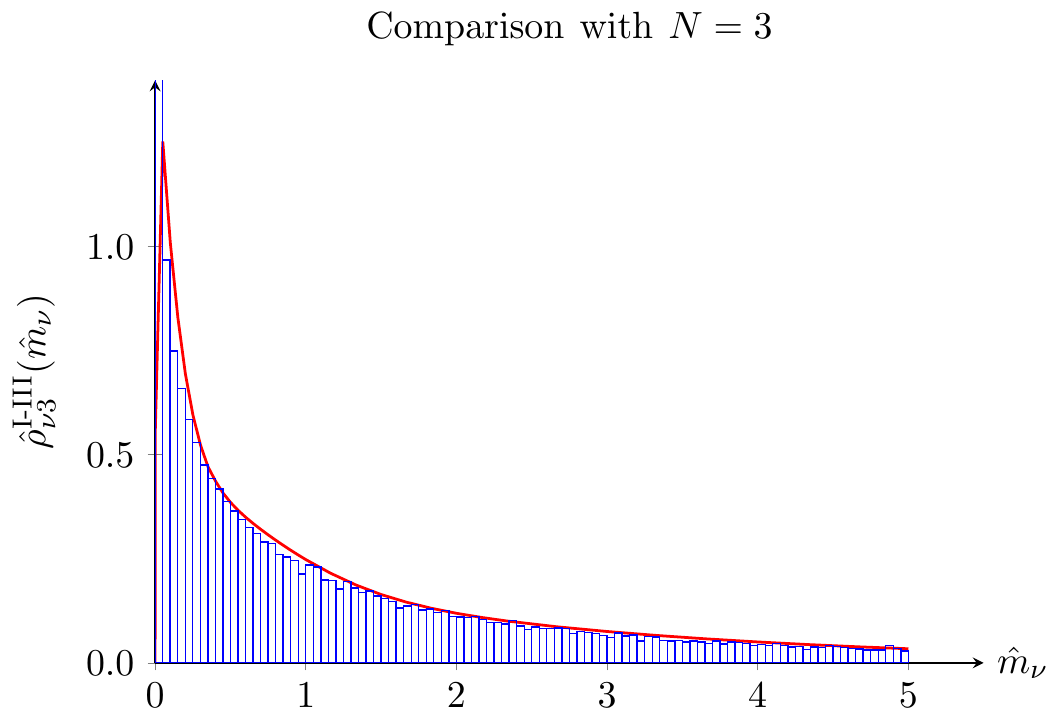}
}
\caption{Comparison between the correlation function at $N=2$ and $N=3$ (red curves) and the level density at large $N$ (histogram with $N=60$) for the type I-III complex seesaw ensemble.  The red curves correspond to the analytic result for these specific values of $N$ while the histograms correspond to numerical results (with $10^3$ dimensionless light neutrino mass matrices generated).  The left and right panel show the comparison with $N=2$ and $N=3$ respectively.  There is no ordering of the singular values.}
\label{FigPDFlargeN23}
\end{figure}

From figure~\ref{FigPDFlargeN23}, one can see that the agreement between analytical and numerical results becomes surprisingly good as $N$ reaches 3.  However, there is \textit{a priori} no clue as to why the correlation functions for finite and small $N$ are able to reproduce with great precision the level density at large $N$ since they are independent quantities.  This represents the first clear indication that a proper large $N$ analysis would indeed be a good approximation of the physical case $N=3$, as was previously suggested (without proof or evidence) in the literature.  Thus, there is no doubt that this particular behavior should be investigated further since it motivates the search for an analytical expression (coming from a large $N$ analysis) to better understand the physical case at hand.


\section{Discussion and Conclusion}\label{SConclusion}

In this work the statistical implications of the seesaw ensembles, following the anarchy principle, for the physical case of three neutrinos were obtained.  It is shown that the analytic pdfs computed in \cite{Fortin:2016zyf} are in perfect agreement with the numerical results of randomly-generated light neutrino mass matrices for the complex seesaw ensembles with $N=2$ and $N=3$.  The repulsion between the singular values is stronger in the type I-III seesaw ensemble than in the type II seesaw ensemble, and the strength of the difference between the repulsions of type I-III and type II ensembles increases as $N$ increases.

The loss of correlation between the light neutrino masses and the light neutrino mass eigenstates forbids an investigation of the favored hierarchy pattern (normal or inverted).  However, an analysis of the preferred mass splitting, \textit{i.e.} the preferred ordering of $\hat{m}^{2}_{\text{med}}-\hat{m}^{2}_{\text{min}}$ and $\hat{m}^{2}_{\text{max}}-\hat{m}^{2}_{\text{med}}$, is completed.  The probability test implies that for both seesaw ensembles, the preferred mass splitting is the one associated to normal hierarchy, although any permutation of the mass eigenstates is equally likely.  However, a comparison between ensembles shows that the type I-III seesaw ensemble is only twice as likely as the type II seesaw ensemble to generate the neutrino sector experimental data assuming the preferred normal hierarchy.

For all seesaw mechanisms, the preferred neutrino energy scale is of $\mathcal{O}(10^{-2})\:\text{eV}$, which leads to a scale of new physics similar to the GUT scale when the associated coupling constants are of order one.  Smaller coupling constants can partly lower the new physics scale.

A comparison of the group variable pdf for all seesaw ensembles (the Haar measure) and a flat distribution shows that the seesaw ensemble is only twice as likely as the flat distribution to lead to the neutrino sector experimental data.  One thus concludes that the type I-III seesaw ensemble is marginally favored over the other ensembles, predicting the hierarchy of the neutrino sector to be normal.

Finally, a comparison of the complex type I-III seesaw ensemble level density for $N=3$ and large $N$ shows that the properly-normalized $N=3$ level density is well approximated by the properly-normalized large $N$ level density.  Because of the complexity of the analytic $N=3$ singular value pdf and the link between the large $N$ level density and the physical neutrino sector, it would be interesting to obtain an analytical level density at large $N$.  A step in that direction was made in \cite{Fortin:2016zyf} following the usual Coulomb gas technique, but it was shown there that the resulting level density is wrong.  The authors hope to return to this question in the near future.


\ack{
The authors would like to thank Patrick Desrosiers for useful discussions on random matrix theory.  This work is supported by NSERC.
}


\bibliography{SMNeutrinoRMT}

\end{document}